\documentclass[12pt]{iopart}

\usepackage{cite}
\usepackage[normalem]{ulem} 
\usepackage{soul} 
\usepackage{graphicx}
\usepackage{dcolumn}
\usepackage{bm}
\usepackage{euscript,stmaryrd}
\usepackage{xcolor}
\usepackage[utf8]{inputenc}
\usepackage[T1]{fontenc}
\usepackage{mathptmx}
\usepackage{epsf,amsmath,amssymb,verbatim,color,multirow,pifont,mathrsfs,soul,amsthm, wasysym,bm}
\usepackage{graphicx, upgreek}
\usepackage[us,12hr]{datetime}
\usepackage{xcolor}
\usepackage{arydshln}
\usepackage{hyperref}
\usepackage{arydshln}
\usepackage{graphicx}
\usepackage{mathtools}

\theoremstyle{definition}

\newcommand\redout{\bgroup\markoverwith
{\textcolor{red}{\rule[0.5ex]{2pt}{0.8pt}}}\ULon}

\newmuskip\pFqmuskip

\newcommand*\pFq[6][8]{%
  \begingroup 
  \pFqmuskip=#1mu\relax
  \mathcode`\,=\string"8000
  \begingroup\lccode`\~=`\,
  \lowercase{\endgroup\let~}\pFqcomma
  {}_{#2}F_{#3}{\left[\genfrac..{0pt}{}{#4}{#5};#6\right]}%
  \endgroup
}
\newcommand{\pFqcomma}{\mskip\pFqmuskip}

\newcommand{\braket}[2]{\left \langle #1 \middle| #2 \right \rangle}
\newcommand{\braketmatrix}[3]{\left \langle #1 \middle| #2 \middle| #3 \right \rangle}

\usepackage{fancyhdr}
\usepackage{hyperref}
\hypersetup{
    colorlinks,
    citecolor=blue,
    filecolor=blue,
    linkcolor=blue,
    urlcolor=black
}

\begin{document}

\title[Chimera Dynamics of Generalized Kuramoto-Sakaguchi Oscillators in Two-population Networks]{Chimera Dynamics of Generalized Kuramoto-Sakaguchi Oscillators in Two-population Networks}

\author{Seungjae Lee \& Katharina Krischer}

\address{Physik-Department, Technische Universit\"at M\"unchen, James-Franck-Stra\ss e 1, 85748 Garching, Germany}
\ead{seungjae.lee@tum.de}
\ead{krischer@tum.de}
\vspace{10pt}
\begin{indented}
\item[]\today
\end{indented}

\begin{abstract}
Chimera dynamics, an intriguing phenomenon of coupled oscillators, is characterized by the coexistence of coherence and incoherence, arising from a symmetry-breaking mechanism. Extensive research has been performed in various systems, focusing on a system of Kuramoto-Sakaguchi (KS) phase oscillators. In recent developments, the system has been extended to the so-called generalized Kuramoto model, wherein an oscillator is situated on the surface of an $M$-dimensional unit sphere, rather than being confined to a unit circle. In this paper, we exploit the model introduced in [New. J. Phys. \textbf{16}, 023016 (2014)] where the macroscopic dynamics of the system was studied using the extended Watanabe-Strogatz transformation both for real and complex spaces. Considering two-population networks of the generalized KS oscillators in 2D complex spaces, we demonstrate the existence of chimera states and elucidate different motions of the order parameter vectors depending on the strength of intra-population coupling. Similar to the KS model on the unit circle, stationary and breathing chimeras are observed for comparatively strong intra-population coupling. Here, the breathing chimera changes their motion upon decreasing intra-population coupling strength via a global bifurcation involving the completely incoherent state. Beyond that, the system exhibits periodic alternation of the two order parameters with weaker coupling strength. Moreover, we observe that the chimera state transitions into a componentwise aperiodic dynamics when the coupling strength weakens even further. The aperiodic chimera dynamics emerges due to the breaking of conserved quantities that are preserved in the stationary, breathing and alternating chimera states. We provide a detailed explanation of this scenario in both the thermodynamic limit and for finite-sized ensembles. Furthermore, we note that an ensemble in 4D real spaces demonstrates similar behavior.
\end{abstract}

%
%
%
%
%

\section{\label{sec:intro}Introduction}

Within the framework of the concept `\textit{More is different}'~\cite{anderson1972more} or `\textit{More than the sum}'~\cite{Strogatz2022}, collective behavior of systems of nonlinearly coupled objects has been extensively explored in diverse interdisciplinary fields of science and mathematics, with a particular focus on systems of coupled oscillators~\cite{pikovksy_sync,strogatz_sync}. One notable phenomenon in systems of numerous oscillators is the emergence of chimera states~\cite{Omelchenko_2013,Omelchenko_2018,Panaggio_2015}, which are characterized by the simultaneous presence of coherence and incoherence through the breaking of symmetry. Chimera states were originally observed in a system of coupled oscillators on a ring geometry as a spatiotemporal pattern where the system exhibits both a coherently oscillating part and an incoherently oscillating part along the ring~\cite{kuramoto2002}. Later this coexistence pattern was dubbed a chimera state to highlight its peculiar characteristics~\cite{abrams2004}.

For the exploration of the essential properties of chimera states,  Abrams \textit{et al.} exploited a simplified model where identical Kuramoto-Sakaguchi (KS) phase oscillators are stationed in two-population networks~\cite{abrams_chimera2008}. Note that before their work,  Montbri\'o,  Kurths, and Blasius observed chimera states in two-population networks, however, with heterogeneous KS oscillators~\cite{Kurths_twogroup}. In these studies, the intra- and inter-population connections are considered to be all-to-all, with different coupling strengths where the former is stronger than the latter. This configuration effectively emulates a nonlocal coupling scenario on a ring geometry, and since their introduction extensive research has been conducted on the collective dynamics in two-population networks of coupled oscillators. Thereby, the two-population network was considered with identical~\cite{abrams_chimera2008,abrams_chimera2016} and heterogeneous~\cite{hetero_twogroup,hetero_twogroup2,alternating2} KS oscillators. Also, numerous other variations were taken into account, including phase oscillators under higher-order interaction and planar oscillators~\cite{lee1,Laing_SL2010,Laing_SL2019,pazo_winfree,sym_twogroup,olmi_chaos,Olmi_rotator,Bick_2016_chaotic}. 
In addition, chimera states have been ceaselessly studied by adopting three- and multi-population networks as underlying system topology~\cite{martens_three,martens_three2,lee2,laing_ring,bick_2018,Bick2019_m1,bick2022multi,lee_hetero}.

Investigations of the collective dynamics of sinusoidally coupled oscillators, such as Kuramoto oscillators on the unit circle, have relied heavily on dimension reduction methods. Notably, the Ott-Antonsen (OA) ansatz~\cite{OA1,OA2} and the Watanabe-Strogatz (WS) transformation~\cite{WS_original1,WS_original2,WS_mobius} were extensively employed, including the exploration of chimera states~\cite{pikovsky_WS1,pikovsky_WS2,Laing_OA,Bick2020}. The introduction of the WS or OA ansatz allows for the description of the system's macroscopic dynamics, providing an alternative to the Kuramoto order parameter which is defined from the microscopic phase information of the system. For details, see Sec.~\ref{Sec:review}.

Meanwhile, Kuramoto oscillators, originally defined on the unit circle in $\mathbb{C}^1$ or $\mathbb{R}^2$, have been extended to so-called generalized Kuramoto oscillators~\cite{g_KM2,g_KM7,g_KM8,g_KM9,g_KM10,g_KM11,deaguiar2023numerical}. In these models, each oscillator is represented as a unit vector confined to the surface of a higher-dimensional unit sphere, and their collective dynamics, such as synchronization of the oscillators, has been studied. Furthermore, dimension reduction methods like the OA and the WS ansatz were also developed and employed for the study of the generalized Kuramoto oscillators~\cite{g_KM1,g_KM3,g_KM4,g_KM5,g_KM6,g_KM12,g_KM13,Tanaka_2014}. In this paper, we exploit, in particular, the model suggested by T. Tanaka in Ref.~\cite{Tanaka_2014} where the author proposed the generalized Watanabe-Strogatz transformation as a vector form of a linear fractional transformation. This projection map allows for the investigation of the macroscopic dynamics of the system in the thermodynamic limit, considering both real ($\mathbb{R}^M$) and complex ($\mathbb{C}^M$) spaces. 

The rest of this paper is organized as follows. In Sec.~\ref{Sec:review}, we revisit the generalized Kuramoto model and the generalized WS transformation introduced in Ref.~\cite{Tanaka_2014}, necessary for our study. In Sec.~\ref{sec:BF}, we introduce a suitable coupling matrix that determines the Benjamin-Feir instability point and corresponds to the phase-lag parameter in the standard Kuramoto-Sakaguchi model. Then, in Sec.~\ref{sec:complex_system}, we discuss observable chimera states of the generalized KS oscillators in two-population networks with a particular focus on the 2D complex spaces ($\mathbb{C}^2$). Therein, the chimera states are investigated in the thermodynamic limit, using the WS macroscopic variables, and in the finite-sized ensembles, using the Kuramoto order parameter, respectively. Furthermore, in Sec.~\ref{sec:real_system}, we also provide the scenario of emergence of chimera states in the 4D real spaces ($\mathbb{R}^4$) in the thermodynamic limit as a comparison to the 2D complex spaces. Finally, we summarize our results in Sec.~\ref{sec:Conlcusion}. Some useful information will be given through \ref{append:order_parameter}-\ref{append:derivation2}.

\section{\label{Sec:review}Dimension Reduction Methods}

A system of identical Kuramoto-Sakaguchi phase oscillators is governed by
\begin{flalign}
\dot{\phi}_j(t) &= \omega + 2~\text{Im}\bigg[ g(t) e^{-i\phi_j}\bigg] \notag \\
&=\omega + \frac{2}{N}\sum_{k=1}^{N}\sin(\phi_k-\phi_j-\alpha) \label{eq:KS_phase}
\end{flalign} for $j=1,...,N$. Each oscillator $\phi_j(t) \in \mathbb{T}:=[0,2\pi]$ is defined on a unit circle of $\mathbb{C}^1$, i.e., by the argument of a complex number $e^{i\phi_j(t)}$ with unit modulus. The oscillator is influenced sinusoidally by the mean-field forcing $g(t):=e^{-i\alpha} m(t) \in \mathbb{C}^1$. Here, the Kuramoto order parameter $m(t)$ is defined by
\begin{flalign}
m(t) := \frac{1}{N}\sum_{k=1}^{N}e^{i\phi_k(t)} \in \mathbb{C}^1 \label{eq:Kuramoto_OP}
\end{flalign} which serves as the centroid of the phases on the unit circle\cite{STROGATZ20001}. The phase-lag parameter $\alpha \in [0,2\pi]$ determines the Benjamin-Feir instability point~\cite{kuramoto2002,cgle}, i.e., $\alpha_{\text{BF}}=\frac{\pi}{2}$: for $\alpha < \alpha_{\text{BF}}$, the synchronized state $\{\phi_j(t)=\phi_0 +\Omega t \}_{j=1}^{N}$ is stable whereas it becomes unstable for $\alpha > \alpha_{\text{BF}}$, where $\Omega$ is the common locked frequency. On the other hand, desynchronized states, e.g., a splay state $\{\phi_j(t) = \frac{2\pi}{N}j +\Omega t \}_{j=1}^{N}$, is stable for $\alpha > \alpha_\text{BF}$ whereas it becomes unstable for $\alpha < \alpha_\text{BF}$~\cite{WS_original2}. All the oscillators have the same intrinsic frequency $\omega \in \mathbb{R}$ that can be set to zero due to rotational symmetry. 

To investigate the collective behavior of the system, one can exploit the Watanabe-Strogatz (WS) transformation~\cite{WS_original2,pikovsky_WS1,WS_mobius}, which is a linear fractional transformation~\cite{stein2010complex} that reads
\begin{flalign}
e^{i\phi_j(t)} = M_t(e^{i\theta_j}) := \frac{b(t)+e^{i\theta_j}e^{i\varphi(t)} }{1+\overline{b(t)}e^{i\theta_j}e^{i\varphi(t)} } \label{eq:Mobius_TR}
\end{flalign} where $\varphi(t)\in \mathbb{R}$ and $b(t) \in \mathbb{C}$ are called the WS variables characterizing the macroscopic dynamics of the system. The symbol with an overbar represents the complex conjugate. In particular, the WS variable $b(t)$ quantifies the degree of coherence of the system in a similar manner (although not precisely the same) as the Kuramoto order parameter, and fulfills $|b(t)|<1$ for all $t$~\cite{pikovsky_WS2,WS_mobius}. Here, $\{\theta_j\}_{j=1}^{N}$ is a set of constants of motion obeying three constraints: $\sum_{k=1}^N e^{i\theta_k} =0$ in order for $b(t)=0$ to be consistent with the incoherent state $m(t)=0$, and the third one can be written as $\sum_{k=1}^N\theta_k=0$. They are determined by a given initial condition $\{\phi_j(0)\}_{j=1}^{N}$ (see Ref.~\cite{pikovsky_WS2}). The WS macroscopic variables are governed by $\dot{b} = -\overline{g}b^2+g$ and $\dot{\varphi}=i \overline{g} b-i g \overline{b}$~\cite{WS_mobius}.

In the thermodynamic limit ($N \rightarrow \infty$) with uniform constants of motion, one can approach the Ott-Antonsen (OA) manifold~\cite{OA1,OA2} where the WS variable $b(t)$ exactly coincides with the Kuramoto order parameter $m(t)$~\cite{pikovsky_WS1,pikovsky_WS2}. To see this, we first consider the uniform distribution of constants of motion $d\mu(\theta) = \frac{1}{2\pi}d\theta$ together with the Watanabe-Strogatz transformation $\phi = T(\theta) := -i \log M_t(e^{i\theta})$ for a fixed $t$. This transformation pushes-forward a measure $\mu$ to $T_{*}\mu$ and gives $d(T_{*}\mu)(\phi) = f(\phi)d\phi$ where the phase density function
\begin{flalign}
f(\phi)&:=\frac{1}{2\pi}\partial_{\phi}T^{-1}(\phi) \notag \\
&= \frac{1}{2\pi}\frac{1-|b|^2}{1-2|b|\cos(\phi-\arg b) + |b|^2} \label{eq:Poisson_kernel}
\end{flalign} is the normalized Poisson kernel~\cite{WS_mobius}. This fact reveals that the oscillators' phases are distributed according to the normalized Poisson kernel in the OA manifold~\cite{Laing_OA}. The Poisson kernel has the Kuramoto order parameter given by
\begin{flalign}
m(t) &= \int_\mathbb{T} M_t(e^{i\theta}) d\mu(\theta)= \int_\mathbb{T} e^{i\phi} d(T_{*}\mu)(\phi)   \notag \\
&= \int_\mathbb{T} e^{i\phi}  \frac{1}{2\pi} \frac{1-|b|^2}{1-2|b|\cos(\phi-\arg b) + |b|^2} d\phi \notag \\
&= b(t). \label{eq:order_parameter_and_WS_variable}
\end{flalign} As a result, it becomes possible to examine the macroscopic behavior (i.e, the dynamics of the Kuramoto order parameter) of the system by focusing solely on the dynamics of the WS variable $b(t)$ within the OA manifold~\cite{pikovsky_WS2,WS_mobius,Laing_OA}.

The system of Kuramoto-Sakaguchi oscillators (\ref{eq:KS_phase}) defined on a unit circle in $\mathbb{C}^1$ or $\mathbb{R}^2$ has a sinusoidal form and therefore can be extended to a system of generalized Kuramoto oscillators that are defined on the surface of a unit sphere in $M$-dimensional real and complex spaces~\cite{Tanaka_2014}. To get a brief glimpse of this, we can write Eq.~(\ref{eq:KS_phase}) as
\begin{flalign}
\frac{d}{dt}e^{i\phi_j} &= i \dot{\phi}_je^{i\phi_j} = i\omega e^{i\phi_j} + 2ie^{i\phi_j} \text{Im}\bigg[g(t)e^{-i\phi_j}\bigg] \notag \\
&= i\omega e^{i\phi_j} + e^{i\phi_j} \bigg( g(t)e^{-i\phi_j} - \overline{g(t)}e^{i\phi_j} \bigg) \notag \\
&= -e^{i\phi_j}\overline{g(t)} e^{i\phi_j} +i\omega e^{i\phi_j}  + g(t) 
\end{flalign} for $j=1,...,N$. Let us take $e^{i\phi_j(t)} \in \mathbb{C}^1 \mapsto \bm{x}_j(t) \in \mathbb{K}^M$ with a constraint $\bm{x}_j(t)^{\dag}\bm{x}_j(t) = \braket{\bm{x}_j(t)}{\bm{x}_j(t)} = 1$ for all $t$ and $j=1,...,N$ where $\mathbb{K} = \mathbb{R}$ or $\mathbb{C}$ is a ground field, and $\dag$ denotes a Hermitian adjoint~\cite{sakurai1967advanced}. Then, one can treat $\bm{x}_j(t)$ as an oscillator defined on the surface of the unit sphere denoted by $\mathbb{S}^{M} := \{\bm{x} \in \mathbb{K}^M | \bm{x}^\dag \bm{x} =1 \}$ either for $\mathbb{K}=\mathbb{R}$ or $\mathbb{C}$. In this consideration, the oscillators can be assumed to follow
\begin{flalign}
\dot{\bm{x}}_j = -\bm{x}_j \bm{g}^{\dag} \bm{x}_j +\Omega \bm{x}_j + \bm{g}
\end{flalign} for $j=1,...,N$. Here, the mean-field forcing $\bm{g}(t) \in \mathbb{K}^M$ can be any arbitrary vector that characterizes an external forcing function or a global coupling of the system. In this paper, we consider the mean-field forcing $\bm{g}(t)$ which depends only on the Kuramoto order parameter (see Sec.~\ref{sec:BF} for details). The Kuramoto order parameter serves as the center of mass of the oscillators on $\mathbb{S}^{M}$, which is defined by
\begin{flalign}
\bm{m}(t):=\frac{1}{N}\sum_{k=1}^{N}\bm{x}_k(t) \in \mathbb{K}^M. \label{eq:COM}
\end{flalign} Furthermore, $\Omega \in \mathbb{K}^{M\times M}$ with $\Omega^\dag = -\Omega$ is an anti-hermitian natural frequency matrix that corresponds to $i\omega$ in the above equation. Throughout this paper,  we will set $\bm{\Omega}$ to zero since we only consider identical oscillators. This model and other variants of vector-formed oscillators, i.e., generalized Kuramoto models have been so far investigated for a variety of collective dynamics~\cite{g_KM3,g_KM4,g_KM5,g_KM7,g_KM8,g_KM9,g_KM10,g_KM11}. In particular, the macroscopic dynamics of the generalized Kuramoto oscillator system can be investigated based on the generalized WS transformation introduced in Ref.~\cite{Tanaka_2014}.

Below, we briefly revisit the main focus of Ref.~\cite{Tanaka_2014} that is necessary for our study. Therein, the higher-dimensional WS transformation was introduced as a vector form of a linear fractional transformation 
\begin{flalign}
\bm{x}_j(t) = M_t(\bm{x}_{0,j}) := \frac{\bm{A} \bm{x}_{0,j} + \bm{b} }{\bm{b}^{\dag}\bm{A}\bm{x}_{0,j}+1} \label{eq:Tanaka_WS}
\end{flalign} where $\bm{A}(t) \in \mathbb{K}^{M \times M}$ and $\bm{b}(t) \in \mathbb{K}^M$ are the WS variables that determine the macroscopic dynamics of the system. In this context, the initial conditions $\bm{x}_j(0) =: \bm{x}_{0,j}$ act as the constants of motion. To describe the incoherent state consistently, i.e., $\bm{m}(t)=\bm{b}(t)=0$, the initial conditions should satisfy $\frac{1}{N}\sum_{k=1}^{N}\bm{x}_{0,k}=0$. This holds, e.g., if they are uniformly distributed on $\mathbb{S}^{M}$. Under the constraint $\braket{\bm{x}_j}{\bm{x}_j}=1$ and with uniform initial conditions, the WS variables are governed by
\begin{flalign}
\dot{\bm{A}} &= \bm{g}\bm{b}^\dag \bm{A}-\bm{g}^\dag \bm{b}\bm{A} \notag \\
\dot{\bm{b}} &= -\bm{b}\bm{g}^\dag \bm{b}+ \bm{g}. \label{eq:WS_dynamics_Tanaka}
\end{flalign} Furthermore, one can show that $\bm{A} = \bm{H}^{1/2}\bm{U}$ (polar decomposition) where $\bm{U} \in \mathbb{K}^{M\times M}$ is a unitary matrix and $\bm{H}^{1/2} = \bm{V} \Sigma^{1/2} \bm{V}^\dag$ with $\Sigma^{1/2} := (\sqrt{\Sigma_{ij}})$. Here, $\bm{V} \Sigma \bm{V}^\dag$ is a singular value decomposition of $\bm{H} := (1-|\bm{b}|^2)I_M + \bm{b}\bm{b}^\dag$ where $I_M \in \mathbb{R}^{M\times M}$ is the identity matrix. Thus it follows that $\bm{H}^{1/2}\bm{b}=\bm{b}$. 

In the thermodynamic limit, without loss of generality, it is possible to set $\bm{U} = I_M$ as long as $\bm{x}_{0,j}$ are uniformly distributed on $\mathbb{S}^{M}$. Then, as T. Tanaka showed~\cite{Tanaka_2014}, the Kuramoto order parameter can also be expressed in terms of the WS (or OA) variables as
\begin{flalign}
\bm{m}(t) &= \int_{|\bm{x}_0|=1} M_t(\bm{x}_0) d\mu(\bm{x}_0) = \frac{1}{S_M}\int_{|\bm{x}_0|=1} \bm{x}(t) d\bm{x}_0 \notag \\
&= \frac{M-1}{M}  \, _2F_1(\frac{1}{2},1;\frac{M+2}{2};|\bm{b}|^2)\bm{b}(t) \notag \\
& = h(|\bm{b}|^2,M) \bm{b}(t)
\label{eq:order_real}
\end{flalign} for $\mathbb{K}=\mathbb{R}$. Here, $\, _2F_1$ is the ordinary hypergeometric function~\cite{hassani2013mathematical}, $S_M:=\frac{2\pi^{M/2}}{\Gamma(M/2)}$ is the surface area of the $(M-1)$-sphere, and $h(|\bm{b}|^2,M): =\frac{M-1}{M}  \, _2F_1(\frac{1}{2},1;\frac{M+2}{2};|\bm{b}|^2) $. For $\mathbb{K}=\mathbb{C}$, the Kuramoto order parameter is given by
\begin{flalign}
\bm{m}(t) &= \frac{1}{S_{2M}}\int_{|\bm{x}_0|=1} \bm{x}(t) d\bm{x}_0 = \bm{b}(t) \label{eq:order_complex}
\end{flalign} which means the Kuramoto order parameter exactly coincides with the WS variable $\bm{b}(t)$. For the details, see Ref.~\cite{Tanaka_2014} and ~\ref{append:order_parameter}.

\section{\label{sec:BF}Coupling Matrix: Benjamin-Feir Instability Point}

To explore chimera states in a system of generalized Kuramoto-Sakaguchi oscillators in two-population networks, we introduce a suitable coupling matrix in this section, which corresponds to the phase-lag parameter $\alpha$ in Eq.~(\ref{eq:KS_phase}). Moreover, the Benjamin-Feir (BF) instability point is obtained for this coupling matrix. In the remainder of this paper, we consider the system below the BF point, in order to follow the previous observations of chimeras in two-population networks~~\cite{abrams_chimera2008,abrams_chimera2016,Laing_SL2010,Laing_SL2019,lee1}.

In line with Eq.~(\ref{eq:KS_phase}), we define the forcing function as $\bm{g}(t) = \bm{K} \bm{m}(t)$ where $\bm{K} \in \mathbb{K}^{M\times M}$ is a coupling matrix and $\bm{m}(t)$ is the Kuramoto order parameter in Eq.~(\ref{eq:COM}). Then, the microscopic dynamics of the oscillators are given by~\cite{Tanaka_2014}
\begin{flalign}
\dot{\bm{x}}_j &= -\bm{x}_j \bm{g}^\dag \bm{x}_j +\bm{g} \notag \\
&=-\bm{x}_j\bm{m}^\dag \bm{K}^\dag \bm{x}_j + \bm{K}\bm{m} = \frac{1}{N}\sum_{k=1}^{N} \bigg( \bm{K} \bm{x}_k - (\bm{x}_k^\dag \bm{K}^\dag \bm{x}_j ) \bm{x}_j  \bigg)
\end{flalign} for $j=1,...,N$. In Eq.~(\ref{eq:KS_phase}), the phase-lag parameter induces phase rotations of $-\alpha$ to each oscillator's phase on the unit circle. Similarly, we introduce a rotation matrix as the coupling matrix~\cite{g_KM11,g_KM13}. First, we consider the real space $\mathbb{K} = \mathbb{R}$. For a rotation in the real space, we need to distinguish between even and odd dimensional cases~\cite{g_KM4,g_KM9}. For odd $M$, we set the $M$-axis (see \ref{append:order_parameter}) as the rotational axis; then $M^\perp$-plane is isoclinically rotating with the same angle. An example is
\begin{flalign}
\bm{K} = \begin{pmatrix}
 \cos\alpha & -\sin\alpha  & 0 & 0  &0\\
\sin\alpha & \cos\alpha & 0 &0 &0 \\
 0& 0 & \cos\alpha & -\sin\alpha&0 \\
 0 & 0 & \sin\alpha & \cos\alpha &0 \\
 0 &0 &0 &0 &1
\end{pmatrix}
\end{flalign} for $\mathbb{R}^5$. For even $M$, there is no rotational axis and hence we exploit planes of rotation, in particular, an isoclinic rotation with the same rotational angle for each plane. An example is 
\begin{flalign}
\bm{K} = \begin{pmatrix}
 \cos\alpha & -\sin\alpha  & 0 & 0  \\
\sin\alpha & \cos\alpha & 0 &0  \\
 0& 0 & \cos\alpha & -\sin\alpha \\
 0 & 0 & \sin\alpha & \cos\alpha 
\end{pmatrix} 
\end{flalign} for $\mathbb{R}^4$. 

To determine the Benjamin-Feir point, we consider the macroscopic WS dynamics in the thermodynamic limit Eq.~(\ref{eq:WS_dynamics_Tanaka}), but use the notation $\bm{\psi}$ instead of $\bm{b}$ to denote the WS variable for future reference. The WS dynamics is then written as
\begin{flalign}
\dot{\bm{\psi}} &= - \bm{\psi} \bm{g}^\dag \bm{\psi} + \bm{g} \notag \\
&= h(|\bm{\psi}|^2,M)\bigg(-\bm{\psi} \bm{\psi}^\dag \bm{K}^\dag \bm{\psi} + \bm{K} \bm{\psi} \bigg) \label{eq:one_pop_governing}
\end{flalign} where the order parameter $\bm{m}= h(|\bm{\psi}|^2,M) \bm{\psi}$ as in Eq.~(\ref{eq:order_real}). Using Eq.~(\ref{eq:one_pop_governing}), the dynamics of the magnitude of the WS variable is given by
\begin{flalign}
\partial_t |\bm{\psi}|^2 &= \partial_t (\bm{\psi}^\dag \bm{\psi}) \notag \\
&=h(|\bm{\psi}|^2,M) \bigg( (-\bm{\psi}^\dag \bm{K}\bm{\psi}\bm{\psi}^\dag + \bm{\psi}^\dag \bm{K}^\dag)\bm{\psi} +\bm{\psi}^\dag (-\bm{\psi}\bm{\psi}^\dag \bm{K}^\dag \bm{\psi}+\bm{K}\bm{\psi})\bigg) \notag \\
& = 2 h(|\bm{\psi}|^2,M)(1-|\bm{\psi}|^2)\braketmatrix{\bm{\psi}}{\frac{\bm{K}+\bm{K}^\dag}{2}}{\bm{\psi}} \label{eq:one_pop_magnitude_dynamics}
\end{flalign} where $\partial_t := \frac{d}{dt}$. For even $M$, the last term becomes $\braketmatrix{\bm{\psi}}{\frac{\bm{K}+\bm{K}^\dag}{2}}{\bm{\psi}} = \cos\alpha |\bm{\psi}|^2 $. Thus, Equation (\ref{eq:one_pop_magnitude_dynamics}) is simply written as
\begin{flalign}
\partial_t|\bm{\psi}|^2 = 2 h(|\bm{\psi}|^2,M) (1-|\bm{\psi}|^2) |\bm{\psi}|^2 \cos\alpha. 
\end{flalign} There exist two fixed points: $\rho_*=|\bm{\psi}|=0$ representing the completely incoherent state, and $\rho^*=|\bm{\psi}|=1$, indicating the synchronized state. The stability of these fixed points depends on the value of $\alpha$. Specifically, the synchronized state $\rho^*$ is stable while the incoherent state $\rho_*$ is unstable for $\alpha < \frac{\pi}{2}$. Opposite to this, the incoherent state $\rho_*$ is stable whereas the synchronized state $\rho^*$ becomes unstable for $\alpha > \frac{\pi}{2}$. Therefore, the Benjamin-Feir instability occurs at $\alpha_{\text{BF}}=\frac{\pi}{2}$ for even $M$.

On the other hand, for odd $M$, the last term reads as $\braketmatrix{\bm{\psi}}{\frac{\bm{K}+\bm{K}^\dag}{2}}{\bm{\psi}} = |\bm{\psi}|^2\cos\alpha +(1-\cos\alpha)x_M^2$ where $x_M$ is the coordinate of $\bm{\psi}$ along the rotational axis, i.e., the $M$-axis. In this case, the synchronized state exists in two distinct spaces: either on the $M$-axis or on the $M^\perp$-plane. In the former case, the synchronized state is always stable, regardless of the value of $\alpha$ since it is governed by $\partial_t|\bm{\psi}|^2 = 2h(|\bm{\psi}|^2,M) (1-|\bm{\psi}|^2) |\bm{\psi}|^2 $ on the $M$-axis. Note $\bm{\psi}=(0,...,0,x_M)^\top$ and $|\bm{\psi}|=x_M$ along the $M$-axis. For the latter case, the stability of the synchronized state depends on the parameter $\alpha$: Equation (\ref{eq:one_pop_magnitude_dynamics}) on the $M^\perp$-plane can be written as $\partial_t|\bm{\psi}|^2 = 2h(|\bm{\psi}|^2,M) (1-|\bm{\psi}|^2) |\bm{\psi}|^2\cos\alpha$ (here, $x_M=0$ on the $M^\perp$-plane), which reveals the synchronized state is stable for $\alpha <\frac{\pi}{2}$ and unstable for $\alpha > \frac{\pi}{2}$. Consequently, the Benjamin-Feir instability also occurs at $\alpha_\text{BF} = \frac{\pi}{2}$. Note that the synchronized state on the $M^\perp$-plane is always unstable with respect to a perturbation along the $M$-axis. See \ref{append:3d_Benjamin} for the 3D real space. A similar result was reported in Refs.~\cite{g_KM11,g_KM13}.

Moving forward, let us explore the higher-dimensional complex space $\mathbb{C}^M$. In this setting, we choose the coupling matrix as $\bm{K} = e^{-i\alpha}I_M$ for all $M$. Then, the governing equation reads
\begin{flalign}
\dot{\bm{\psi}} &= - \bm{\psi} \bm{g}^\dag \bm{\psi} + \bm{g} \notag \\
&= -\bm{\psi} \bm{\psi}^\dag \bm{K}^\dag \bm{\psi} + \bm{K} \bm{\psi} 
\end{flalign} since the Kuramoto order parameter exactly coincides with the WS variable, i.e., $\bm{m}(t) = \bm{\psi}(t)$ as in Eq.~(\ref{eq:order_complex}). Consequently, the magnitude of the WS variable is determined by
\begin{flalign}
\partial_t|\bm{\psi}|^2 &= 2 (1-|\bm{\psi}|^2)\braketmatrix{\bm{\psi}}{\frac{\bm{K}+\bm{K}^\dag}{2}}{\bm{\psi}} \notag \\
&=2 (1-|\bm{\psi}|^2) |\bm{\psi}|^2\cos\alpha
\end{flalign} Therefore, the synchronized state $|\bm{\psi}|=1$ is stable for $\alpha < \frac{\pi}{2}$ whereas it is unstable for $\alpha>\frac{\pi}{2}$, and thus $\alpha_\text{BF} = \frac{\pi}{2}$.

\section{\label{sec:complex_system}Chimera Dynamics in Two-population Networks for $\mathbb{C}^2$ }

Now, we are prepared to investigate chimera states in two-population networks of identical generalized Kuramoto-Sakaguchi oscillators. The microscopic dynamics of the system is given by
\begin{flalign}
\dot{\bm{x}}^{(1)}_j &= - \bm{x}^{(1)}_j \bm{g}_1^\dag \bm{x}^{(1)}_j + \bm{g}_1 \notag \\
\dot{\bm{x}}^{(2)}_j &= - \bm{x}^{(2)}_j \bm{g}_2^\dag \bm{x}^{(2)}_j + \bm{g}_2 \label{eq:micro_two_pop}
\end{flalign} where the mean-field forcing for each population reads
\begin{flalign}
    \bm{g}_a &= \mu \bm{K} \bm{m}^{(a)} + \nu \bm{K} \bm{m}^{(b)} = \bm{K} \bigg( \frac{\mu}{N}\sum_{k=1}^{N}\bm{x}^{(a)}_k(t) +\frac{\nu}{N}\sum_{k=1}^{N}\bm{x}^{(b)}_k(t)    \bigg)
\end{flalign} and the Kuramoto order parameter of each population is defined by
\begin{equation}
    \bm{m}^{(a)}(t) = \frac{1}{N}\sum_{k=1}^{N}\bm{x}^{(a)}_k(t) \label{eq:order_parameter_complex}
\end{equation} for $(a,b) = (1,2)$ or $(2,1)$. Throughout this work, the inter-population coupling strength is fixed as $\mu=1$ and the intra-population coupling strength $\nu = 1-A$ where $A$ is the control parameter. The rotation matrices for $\mathbb{K}=\mathbb{R}$ in Sec.~\ref{sec:BF} is employed as the coupling matrix while it is $\bm{K} = e^{-i\alpha}I_M$ for $\mathbb{K}=\mathbb{C}$. Note that in this paper, we fix $\alpha = \frac{\pi}{2}-0.005 < \alpha_\text{BF}$.

\subsection{\label{subsec:stationary_breathing}Stationary and Breathing Chimeras}

In the following, we study the chimera states with a particular focus on the 2D complex space $\mathbb{C}^2$. Recall that for the complex space the Kuramoto order parameter $\bm{m}^{(a)}(t)$ is exactly equal to $\bm{\psi}_a(t)$ in the thermodynamic limit. Then, the forcing field is written as $\bm{g}_a = \mu \bm{K} \bm{\psi}_a + \nu \bm{K} \bm{\psi}_b$ where $\bm{K} = e^{-i\alpha}I_M$. Therefore, the WS variable $\bm{\psi}_a$ is governed by
\begin{flalign}
\dot{\bm{\psi}}_a &= - \bm{\psi}_a \bm{g}_a^\dag \bm{\psi}_a + \bm{g}_a \notag \\
&=-\bm{\psi}_a\bigg( \mu \bm{\psi}_a^\dag \bm{K}^\dag +\nu \bm{\psi}_b^\dag \bm{K}^\dag \bigg)\bm{\psi}_a + \mu \bm{K}\bm{\psi}_a +\nu \bm{K} \bm{\psi}_b
\label{eq:WS_complex_two_pop_governing}
\end{flalign} for $(a,b)=(1,2)$ and $(2,1)$ in the thermodynamic limit.

As an initial approach, we examine the dynamics of the magnitude of the order parameter vectors: $|\bm{\psi}_a|=\sqrt{\braket{\bm{\psi}_a}{\bm{\psi}_a}}$ for $a=1,2$. From Eq.~(\ref{eq:WS_complex_two_pop_governing}), the dynamics of the order parameter magnitude reads
\begin{flalign}
\partial_t|\bm{\psi}_1|^2 &= 2(1-|\bm{\psi}_1|^2)\bigg(\mu\cos\alpha |\bm{\psi}_1|^2 \notag + \nu \text{Re}\bigg[e^{-i\alpha}\braket{\bm{\psi}_1}{\bm{\psi}_2}\bigg] \bigg) \notag \\
\partial_t|\bm{\psi}_2|^2 &= 2(1-|\bm{\psi}_2|^2)\bigg(\mu\cos\alpha |\bm{\psi}_2|^2 \notag + \nu \text{Re}\bigg[e^{-i\alpha}\braket{\bm{\psi}_2}{\bm{\psi}_1}\bigg] \bigg) \notag \\
\end{flalign} and the cross term is governed by
\begin{flalign}
\partial_t \braket{\bm{\psi}_1}{\bm{\psi}_2} &= \mu\big((1-|\bm{\psi}_1|^2)e^{-i\alpha} +(1-|\bm{\psi}_2|^2)e^{i\alpha} \big) \braket{\bm{\psi}_1}{\bm{\psi}_2} \notag \\
&~~ ~~+\nu e^{-i\alpha} (|\bm{\psi}_1|^2-\braket{\bm{\psi}_1}{\bm{\psi}_2}^2) + \nu e^{i\alpha}(|\bm{\psi}_2|^2-\braket{\bm{\psi}_1}{\bm{\psi}_2}^2).
\end{flalign} From numerical integration~\cite{Mathematica} of Eq.~(\ref{eq:WS_complex_two_pop_governing}), we find that the cross term can be represented by $\braket{\bm{\psi}_1}{\bm{\psi}_2} = |\bm{\psi}_1||\bm{\psi}_2|e^{i\Theta}$ where $\Theta \in \mathbb{R}$ as explained later in this section. Then, setting $|\bm{\psi}_a| =: \rho_a$ for $a=1,2$ and $\rho_2=1$, we obtain
\begin{flalign}
\partial_t\rho_1 &= \frac{1-\rho_1^2}{\rho_1}\big(\mu\cos\alpha \rho_1^2 + \nu \rho_1 \cos(\Theta-\alpha)\big) \notag \\
\partial_t\Theta &= \mu\sin\alpha (\rho_1^2-1) +\nu \Big( -2\rho_1  \sin(\Theta+\alpha) + \sin(\Theta-\alpha) \big( -\frac{1}{\rho_1}-\rho_1\big) \Big). \label{eq:OA_two_pop}
\end{flalign} Note that the above equations are equivalent to Eq.~(10) in Ref.~\cite{abrams_chimera2008}, i.e., the Ott-Antonsen equation of a system of identical KS oscillators (defined on the unit circle of $\mathbb{C}^1$) in two-population networks. From Eq.~(\ref{eq:OA_two_pop}), we can obtain an overview of some of the observable chimera states.

\begin{figure}[t!]
\centering
\includegraphics[width=0.65\linewidth]{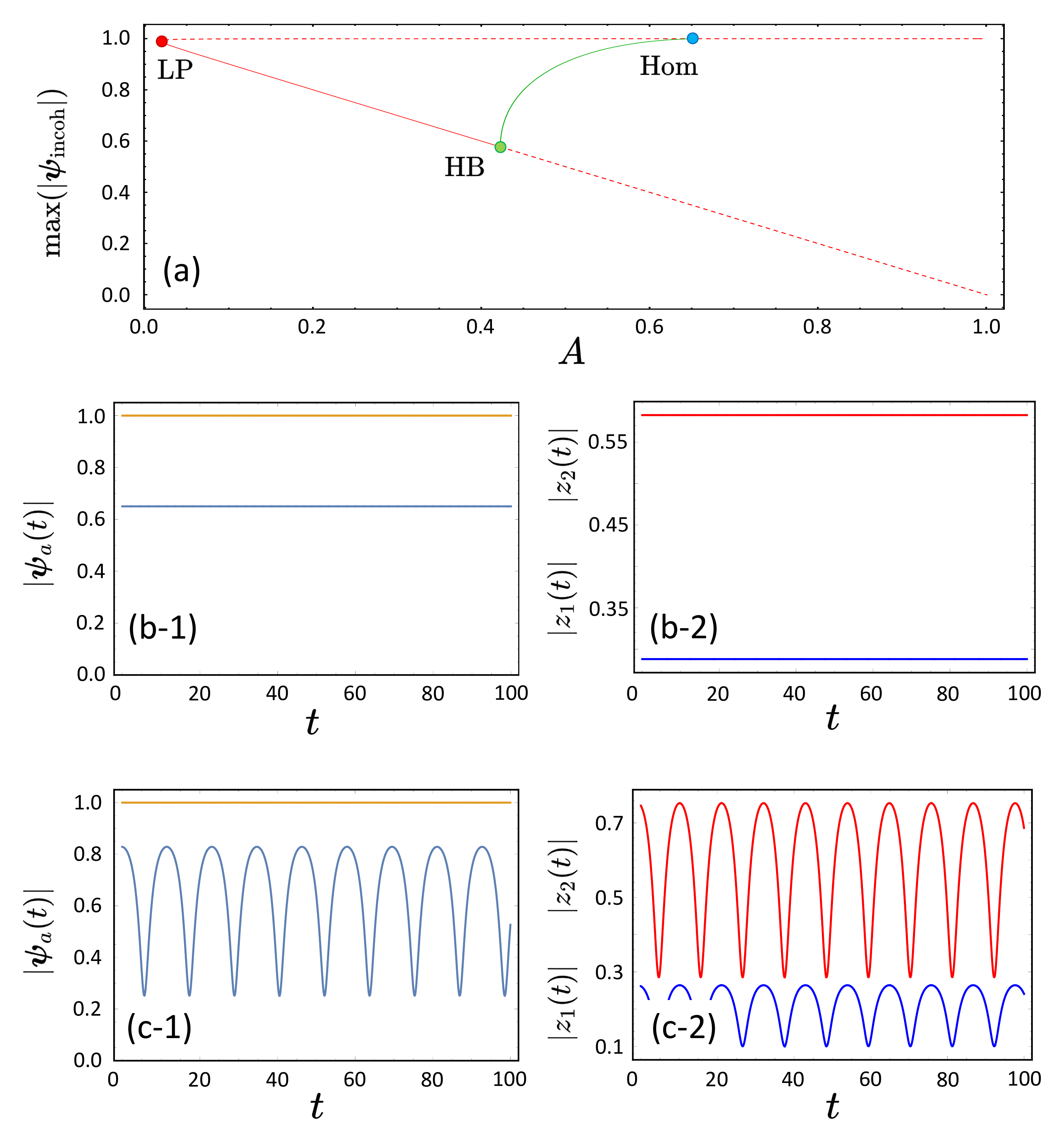}
\caption{(a) A bifurcation diagram of stable (solid curve) and unstable (dashed curve) chimera states from Eq.~(\ref{eq:OA_two_pop}): stationary chimeras (red) and breathing chimera states (green). LP: limiting point bifurcation, HB: supercritical Hopf bifurcation, and Hom: homoclinic bifurcation. Lower panels: Time evolution of the magnitude of the order parameter vectors (the synchronized population $|\bm{\psi}_2|$ in orange and the incoherent population $|\bm{\psi}_1|$ in light blue) in the left column and time evolution of the modulus of the components of the order parameter vector for the incoherent population $\bm{\psi}_1 = (z_1,z_2)^\top$ (the first component $|z_1|$ in red and the second component $|z_2|$ in blue) in the right column. (b) Stationary chimera states for $A=0.35$. (c) Breathing chimera states for $A=0.46$. The presented results are based on data obtained after disregarding the initial transient behavior for $t>10^5$. } 
\label{Fig:bifurcation_complex}
\end{figure}

In Fig.~\ref{Fig:bifurcation_complex} (a), a bifurcation diagram of chimera states is depicted as the parameter $A$ varies. Stable (red solid) and unstable (red dashed) stationary chimera states are born/annihilated at a limiting point (LP) bifurcation for a rather strong intra-population coupling strength. The stable stationary chimera state undergoes a supercritical Hopf bifurcation (HB) at $A=A_\text{HB}$, and a breathing chimera state (green solid) emerges as a stable limit-cycle solution. Integrating Eq.~(\ref{eq:WS_complex_two_pop_governing}) numerically, we find stationary and breathing chimera states for the parameter values according to the bifurcation diagram depicted in Fig.~\ref{Fig:bifurcation_complex} (a). For simplicity, we assume that the first population is incoherent and the second population is synchronized, i.e., $|\bm{\psi}_1(t)|<1$ and $|\bm{\psi}_2(t)|=1$. Also, we use the following notations to denote components of each order parameter vector: $\bm{\psi}_1(t) = (z_1(t),z_2(t))^\top$ and $\bm{\psi}_2(t) = (w_1(t),w_2(t))^\top$ where $z_i(t), w_i(t) \in \mathbb{C}$ and $\theta_i(t) = \arg z_i(t) \in \mathbb{R}$ and $\phi_i(t) =\arg w_i(t) \in \mathbb{R}$ for $i=1,2$. 

In Fig.~\ref{Fig:bifurcation_complex} (b), the time evolution of the magnitude of the order parameter vectors for a stationary chimera state is depicted together with the time evolution of $|z_1(t)|$ and $|z_2(t)|$. Both $|\bm{\psi}_a|$ and the components $|z_i(t)|$ appear as a fixed point solution. Note that for the synchronized population, $|w_i(t)|$ also shows stationary motion. Furthermore, Equation (\ref{eq:WS_complex_two_pop_governing}) is invariant under a unitary transformation defined by
\begin{flalign}
\bm{u} = \begin{pmatrix}
 e^{-i\theta_1}&0  \\
 0 & e^{-i\theta_2} 
\end{pmatrix} \label{eq:unitary_tr}
\end{flalign} which means $F_a(\bm{u} \bm{\psi}_a) = \bm{u} F_a(\bm{\psi}_a)$ where $F_a(\bm{\psi}_a):= - \bm{\psi}_a \bm{g}_a^\dag \bm{\psi}_a + \bm{g}_a$ for $a=1,2$. This unitary transformation is a continuous symmetry corresponding to the phase shift invariance of the KS model on the unit circle. Therefore, the phase difference $\Theta_i(t):= \phi_i(t)-\theta_i(t) $ for $i=1,2$ also shows a stationary motion as a function of time (not shown here). For larger $A$, Figure~\ref{Fig:bifurcation_complex} (c) displays the time evolution of $|\bm{\psi}_a|$ and $|z_i|$ for a given $A \in (A_\text{HB},A_\text{Hom})$. Each variable exhibits a periodic motion as a function of time, and the stable breathing chimera appears indeed as a limit-cycle solution after a supercritical Hopf bifurcation. Further increasing $A$, the breathing chimera state disappears via a homoclinic bifurcation (Hom) as the period of it approaches infinity.

\subsection{\label{subsec:antiphase_chaos}Alternating and Aperiodic Chimeras}

\begin{figure}[t!]
\centering
\includegraphics[width=0.65\linewidth]{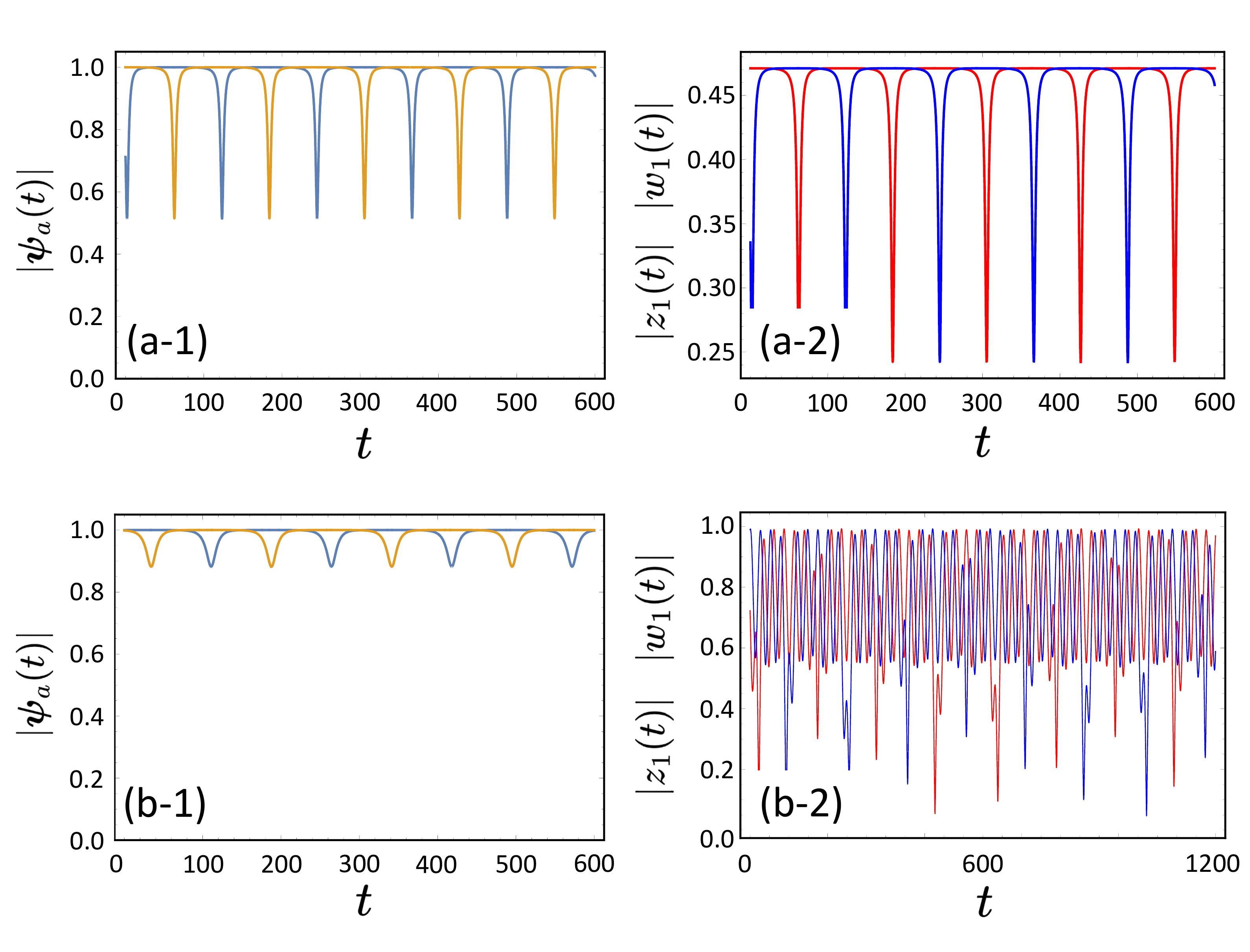}
\caption{ Left column: Time evolution of the magnitude of the order parameters: $|\bm{\psi}_1(t)|$ (blue light) and $|\bm{\psi}_2(t)|$ (orange). Right column: Time evolution of the first component of the order parameter vectors: $|z_1(t)|$ (red) and $|w_1(t)|$ (blue). (a) Periodic alternating chimera at $A=0.75$. (b) Aperiodic alternating chimera at $A=0.86$. The presented results are based on data obtained after disregarding the initial transient behavior for $t>10^5$. } 
\label{Fig:antiphase_chaotic}
\end{figure}

\begin{figure}[t!]
\centering
\includegraphics[width=0.65\linewidth]{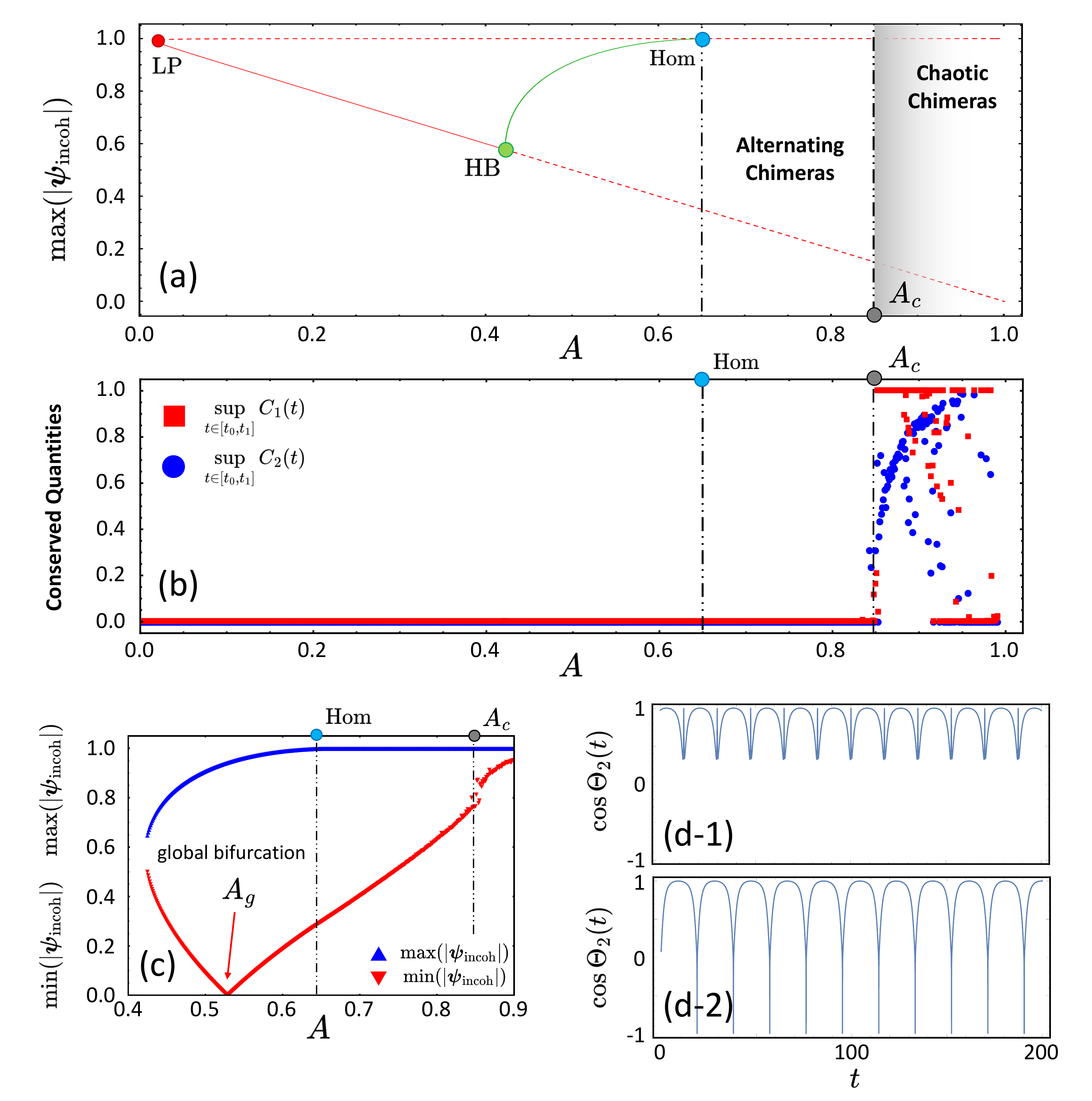}
\caption{ (a) A re-plot of the bifurcation diagram in Fig.~\ref{Fig:bifurcation_complex} (a) with the periodic alternating chimeras and the chaotic chimeras. (b) Conserved quantities as a function of the parameter $A$: $t_0 = 20,000$ and $t_1 =10^5$. (c) Maximum  (blue) and minimum (red) values of the magnitude of the order parameter vectors as a function of the parameter $A$. (d) Time evolution of $\cos\Theta_2(t)$ before (upper, $A=0.52$) and after (lower, $A=0.53$) the global bifurcation. $A_c$ denotes a parameter point from which on the chaotic chimera emerges.} 
\label{Fig:symmetry_broken}
\end{figure}

Thus far, the scenario of the emergence of chimera states for $\mathbb{C}^2$ bears a strong resemblance to that observed in the system of identical KS oscillators on the unit circle of $\mathbb{C}^1$ (e.g., see Refs.~\cite{abrams_chimera2008,abrams_chimera2016}). Nevertheless, in the two-dimensional space $\mathbb{C}^2$, the order parameter consists of two complex components, providing additional complexities beyond this. Let us start with the parameter regime after the homoclinic bifurcation. In Fig.~\ref{Fig:antiphase_chaotic} (a), a chimera state is shown for a given $A>A_\text{Hom}$. This chimera state can be characterized by a periodic alternation between the two order parameter vectors. The magnitude of the order parameter vectors (Fig.~\ref{Fig:antiphase_chaotic} (a-1)) seemingly satisfy $|\bm{\psi}_1(t)| = |\bm{\psi}_2(t-\frac{T}{2})|$ within the limits of our numerical capabilities where $T$ denotes their period. Likewise, each component of the order parameter vectors (Fig.~\ref{Fig:antiphase_chaotic} (a-2)) displays $|z_i(t)| = |w_i(t-\frac{T}{2})|$ for $i=1,2$. Such an alternating chimera motion was reported previously for a system of heterogeneous Kuramoto-Sakaguchi oscillators in two-population networks after the homoclinic bifurcation in which the breathing chimera state disappeared~\cite{alternating2,hetero_twogroup2}. A similar motion was also reported in a system of identical KS oscillators in three-population networks, which as well was born near a homoclinic bifurcation~\cite{lee2}. Together the three observations suggest that the periodic alternating chimera dynamics is somehow linked to the breathing chimera states through the homoclinic bifurcation (see Fig.~\ref{Fig:symmetry_broken} (c)).

\begin{figure}[t!]
\centering
\includegraphics[width=0.65\linewidth]{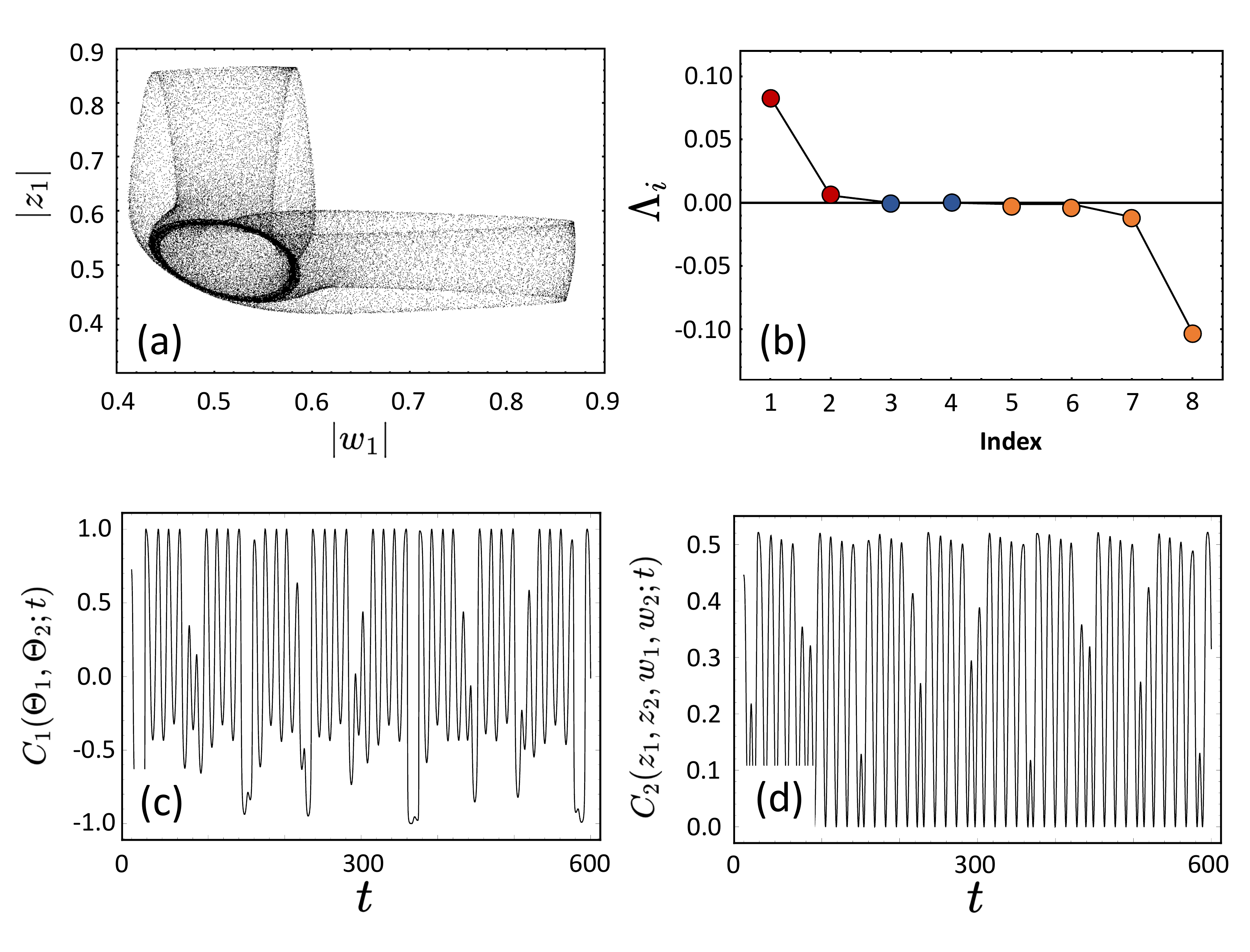}
\caption{ (a) Poincar\'e map of the first components of the order parameters: $|z_1|$ vs. $|w_1|$ measured on the section defined by $\text{Re}[z_2] \equiv 0$. (b) Lyapunov exponents measured along the reference trajectory in (a): Positive exponents (red), zero exponents (blue) and negative exponents (orange). (c-d) Time evolution of the broken conserved quantities of the trajectory in (a) (see main text). The parameter is $A=0.86$ and the transient behavior was discarded ($t>10^5$). } 
\label{Fig:chaotic_detail}
\end{figure}

Following $A$ further, we observe a seemingly similar motion to the alternating chimera dynamics in terms of the magnitude of the order parameters. In Fig.~\ref{Fig:antiphase_chaotic} (b), an example trajectory is shown. The magnitude of the order parameters (Fig.~\ref{Fig:antiphase_chaotic} (b-1)) seems similar to that of the periodic alternating chimera dynamics except for $\text{min}_{t>0}|\bm{\psi}_a(t)|$. However, the componentwise dynamics shows entirely different characteristics. In Fig.~\ref{Fig:antiphase_chaotic} (b-2), we plot an aperiodic time evolution of the first components of the order parameter for both populations. To investigate the aperiodic chimera dynamics in more detail, in Fig.~\ref{Fig:chaotic_detail} (a), the Poincar\'e map of the trajectory is plotted in the section defined by $\text{Re}[z_2] \equiv 0$. The aperiodic dynamics depicted in Fig.~\ref{Fig:antiphase_chaotic} (b-2) displays scattered points on the Poincar\'e section, as anticipated for the aperiodic motion. This conjecture is further supported by the Lyapunov exponents along the reference trajectory in the phase space, following the chimera dynamics in Fig.~\ref{Fig:chaotic_detail} (a)~\cite{pikovsky_LE,CLV1,CLV2}. In Fig.~\ref{Fig:chaotic_detail} (b) it can be seen that there are two positive Lyapunov exponents that indicate a sensitive dependence of the trajectory on initial conditions. Also, there are two zero Lyapunov exponents arising from the two continuous symmetries: time shift (an autonomous equation) and the unitary transformation in Eq.~(\ref{eq:unitary_tr}) corresponding to a phase shift invariance. All the other Lyapunov exponents are negative.

Now, we elucidate how this aperiodic chimera state emerges from the periodic alternating chimera state as $A$ is increased. In Fig.~\ref{Fig:symmetry_broken} (a), the bifurcation diagram is replotted with parameter values indicating where the periodic alternating ($A_\text{Hom}$) and aperiodic ($A_c$) chimera states emerge, respectively. For $A <A_c$, we numerically obtain two conserved quantities along the chimera trajectory in phase space. As explained before, the system is invariant under the unitary transformation in Eq.~(\ref{eq:unitary_tr}). Using this transformation, we can define the phase difference of each component between the two order parameter vectors: $\Theta_i(t):=\phi_i(t)-\theta_i(t)$ for $i=1,2$. The first conserved quantity is given by
\begin{flalign}
C_1(\Theta_1,\Theta_2;t):=\sin(\Theta_1(t)-\Theta_2(t)). \label{eq:sym1}
\end{flalign} For this quantity, $C_1(t)=0$ holds for all $t$ along the chimera trajectory for $A<A_c$, i.e., the stationary, breathing and periodic alternating chimera states all preserve $C_1(t)=0$ along each trajectory. Furthermore, the numerical integration of Eq.~(\ref{eq:WS_complex_two_pop_governing}) for $A<A_c$ confirms the relation between the cross term and the magnitude of the order parameter vectors: $\braket{\bm{\psi}_1}{\bm{\psi}_2} = |\bm{\psi}_1||\bm{\psi}_2|e^{i\Theta}$ where $\Theta:=\Theta_1=\Theta_2$ for $C_1=0$. This relation leads to the second conserved quantity
\begin{flalign}
C_2(z_1,z_2,w_1,w_2;t) :=\big(|z_1(t)||w_2(t)|-|z_2(t)||w_1(t)|\big)^2 \label{eq:sym2}
\end{flalign} whereby $C_2(t)=0$ for all $t$ along the chimera trajectory as long as $A<A_c$.

However, we find that for $A>A_c$, $C_1$ and $C_2$ are not conserved any more along the chimera trajectory. In Fig.~\ref{Fig:chaotic_detail} (c-d), the time evolution of $C_1(t)$ and $C_2(t)$ is depicted for the chimera trajectory in Fig.~\ref{Fig:antiphase_chaotic} (b). Both quantities display irregular deviation from zero as a function of time. Once again, note that for $A<A_c$, $C_1(t)=0$ and $C_2(t)=0$ for all $t$ along the chimera orbit. To see this clearly, we numerically measure $\sup_{t\in[t_0,t_1]}|C_1(t)|$ and $\sup_{t\in[t_0,t_1]}|C_2(t)|$ with respect to the parameter $A$ where $t_0 = 2\times 10^4$ and $t_1 =10^5$. Figure~\ref{Fig:symmetry_broken} (b) confirms that for $A<A_c$ where the stationary, breathing and periodic alternating chimeras are observed, the two quantities remain zero. However, from $A_c$ on, the conserved quantities are broken and begin to show irregular time-evolution along the given chimera trajectory.

Another notable observation is that the breathing chimera state undergoes a global bifurcation involving the completely incoherent state, i.e., $|\bm{\psi}_1|=0$~\cite{hetero_twogroup2}. In Fig.~\ref{Fig:symmetry_broken} (c), the maximum and minimum values of the magnitude of the order parameter vector are depicted for the incoherent population. For small $A$, $\text{min}(|\bm{\psi}_\text{incoh}|)$ is decreasing continuously as $A$ increases up to $A=A_g$. Then, it touches zero value where the completely incoherent state is located. From that point on, the minimum value of the order parameter magnitude exhibits a continuous increase, eventually connecting seamlessly to that of the periodic alternating chimera state for $A>A_\text{Hom}$. Further increasing $A$, we reach $A_c$ where the aperiodic chimera emerges and $\text{min}(|\bm{\psi}_\text{incoh}|)$ undergoes a discontinuous change to higher values. At the global bifurcation $A_g$, the motion of the breathing chimera changes. Figure.~\ref{Fig:symmetry_broken} (d) shows the time evolution of $\cos\Theta(t)$ for the breathing chimera state before (d-1) and after (d-2) the global bifurcation. For $A<A_g$, the phase variable $\Theta(t)$ of the breathing chimera states evolves within some interval smaller than $\mathbb{T}:=[0,2\pi]$. After touching the incoherent state for $A>A_g$, the phase $\Theta(t)$ of the breathing chimera states monotonically increases as a function of time thereby sweeping all $\mathbb{T}$. Thus, at the global bifurcation the motion of the order parameter changes from libration to rotation.

\subsection{\label{subsec:microscopic_complex}Chimera States in the Microscopic Dynamics}

In this section, we investigate the microscopic dynamics of the system of generalized Kuramoto-Sakaguchi oscillators in two-population networks. Here, we directly perform numerical integrations of Eq.~(\ref{eq:micro_two_pop}) for $\mathbb{C}^2$.

\begin{figure}[t!]
\centering
\includegraphics[width=0.65\linewidth]{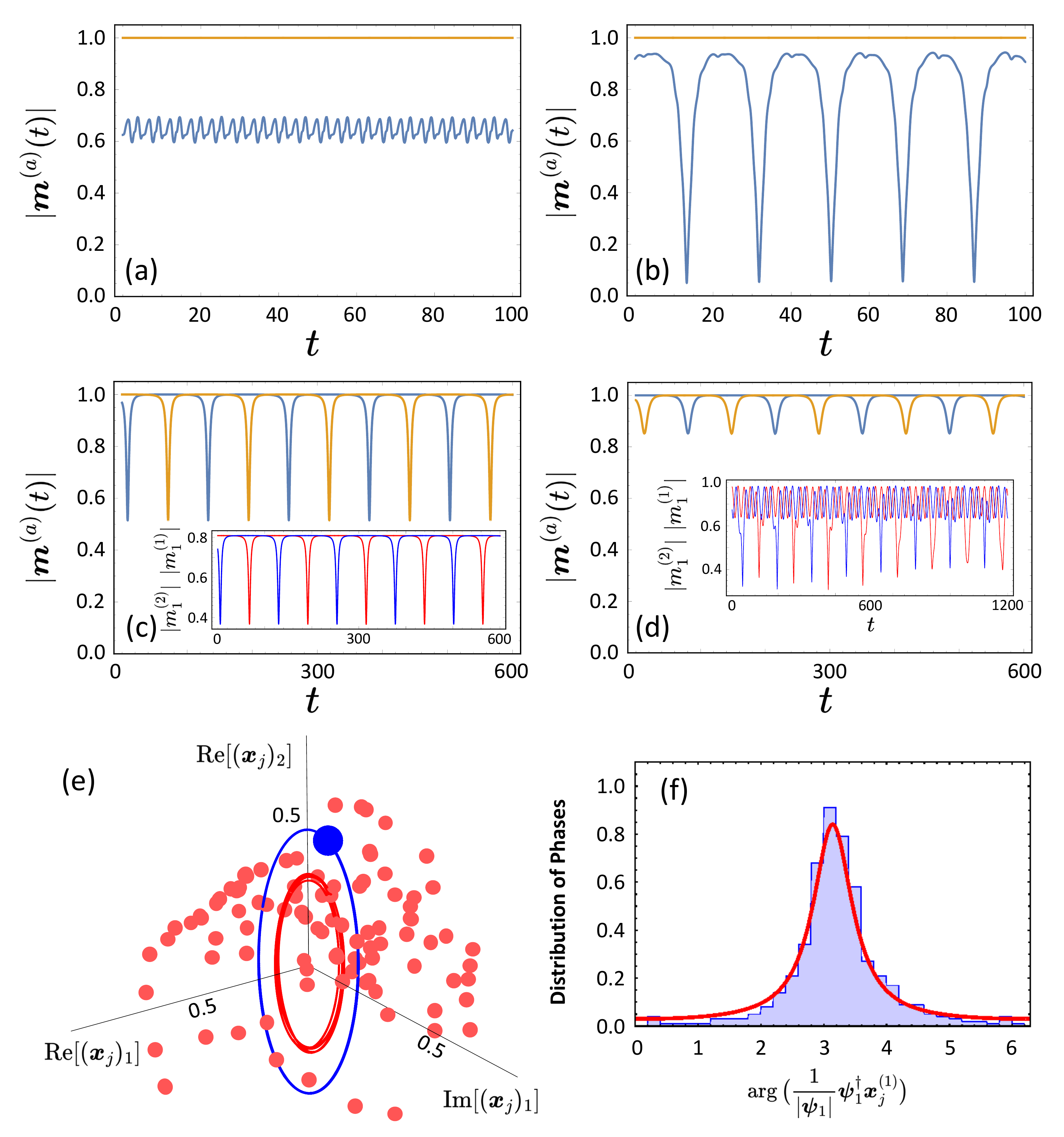}
\caption{(a-d) Time evolution of the magnitude of the order parameter vectors defined in Eq.~(\ref{eq:order_parameter_complex}) from solving Eq.~(\ref{eq:micro_two_pop}) with $N=30$: the first population $|\bm{m}^{(1)}(t)|$ (light blue) and the second population $|\bm{m}^{(2)}(t)|$ (orange). (a) The stationary chimera state with $A=0.35$. (b) The breathing chimera state with $A=0.55$. (c) The periodic alternating chimera state with $A=0.75$. (d) The chimera state of the broken conserved quantities with $A=0.86$. Insets of (c-d): Time evolution of $\bm{m}^{(1)}_1(t)$ (blue) and $\bm{m}^{(2)}_1(t)$ (red). (e) Snapshot of the oscillators with $N=100$. Blue dot: synchronized oscillators $\bm{x}^{(2)}_j$. Red dots: incoherent oscillators $\bm{x}^{(1)}_j$. Blue curve: $\bm{m}^{(2)}(t)$. Red curve: $\bm{m}^{(1)}(t)$. (f) The histogram of the distribution of the phases $ \big\{ \arg\big( \frac{1}{|\bm{\psi}_1|} \bm{\psi}_1^\dag \bm{x}_j^{(1)} \big) \big\}_{j=1}^{N}$ for $N=500$ and $A=0.35$. Red curve indicates the normalized Poisson kernel in Eq.~(\ref{eq:normalized_Poisson}). The presented results are based on data obtained after disregarding the initial transient behavior for $t>10^4$. } 
\label{Fig:micro_complex}
\end{figure}

In Fig.~\ref{Fig:micro_complex} (a-d), the time evolution is depicted for the magnitude of the Kuramoto order parameter $|\bm{m}^{(a)}(t)|$ defined in Eq.~(\ref{eq:order_parameter_complex}) for 30 oscillators in each population $a=1,2$. All the results are obtained from random initial conditions of $\bm{x}^{(a)}_j$ satisfying $\braket{\bm{x}^{(a)}_j(0)}{\bm{x}^{(a)}_j(0)}=1$ for $j=1,...,N$ and $a=1,2$. For a given $A \in (A_\text{LP},A_\text{HB})$, the microscopic dynamics shows a motion of the order parameter vector corresponding to the stationary chimera state (Fig.~\ref{Fig:micro_complex} (a)). Due to the finite-size effect and random initial conditions, we observe a fluctuation in the magnitude of the Kuramoto order parameter. Likewise, in Fig.~\ref{Fig:micro_complex} (b), for $A \in (A_\text{HB},A_\text{Hom})$, the order parameter obtained from the breathing chimera state exhibits small fluctuation superimposed to a simple periodic variation. Furthermore, for $A \in (A_\text{Hom},A_c)$, the microscopic dynamics as well forms alternating chimeras, manifesting themselves in alternating motion of the magnitude and also of each component of the order parameter vectors between two populations (see Inset of Fig.~\ref{Fig:micro_complex} (c)). In the microscopic dynamics, we numerically found that the chimera states of these three types also possess conserved quantities along the trajectory. More precisely, $C_1(t)=0$ and $C_2(t)=0$ with
\begin{flalign}
C_1(t) &:= \sin(\Theta_1(t)-\Theta_2(t)) \notag \\
C_2(t) &:= \big(|m^{(1)}_1(t)||m^{(2)}_2(t)|-|m^{(2)}_2(t)||m^{(1)}_1(t)|\big)^2 
\end{flalign} Here, $\bm{m}^{(1)} = (m^{(1)}_1,m^{(1)}_2)^\top \in \mathbb{C}^2$ and $\bm{m}^{(2)} = (m^{(2)}_1,m^{(2)}_2)^\top \in \mathbb{C}^2$ are the Kuramto order parameters in Eq.~(\ref{eq:order_parameter_complex}), and $\Theta_i = \arg m^{(1)}_i - \arg m^{(2)}_i$ for $i=1,2$. As in the thermodynamic limit, increasing the parameter to be $A >A_c$, $C_1$ and $C_2$ become time-dependent. In Fig.~\ref{Fig:micro_complex} (d), the time evolution of the order parameters is depicted for an example trajectory for $A>A_c$ where the conserved quantities are broken. Also, the components of the order parameter vectors do not show the periodic alternation between two populations but rather they exhibit aperiodic dynamics (see Inset of Fig.~\ref{Fig:micro_complex} (d)). Consequently, our observation of chimera states in the thermodynamic limit can be also verified in the ensembles of a finite number of oscillators. In Fig.~\ref{Fig:micro_complex} (e), a snapshot of a stationary chimera state in a system of 100 oscillators is shown, obtained from Eq.~(\ref{eq:micro_two_pop}). The synchronized oscillators (blue dots) behave alike altogether along the trajectory of the order parameter $\bm{m}^{(2)}(t)$ (blue curve). The incoherent oscillators (red dots) are distributed in the phase space and the order parameter $\bm{m}^{(1)}(t)$ (red curve) has lower magnitude than $\bm{m}^{(2)}(t)$.

Concerning the distribution of the oscillators in the incoherent population, one can at least study the distribution of the incoherent oscillators along the direction of $\bm{\psi}_1$ as follows. Here, we assume the stationary chimera state for a given $A \in (A_\text{LP},A_\text{HB})$ for which $\bm{\psi}_1$ is saturated at a stationary value after a transient behavior. First, define an angular variables of the incoherent oscillators along the order parameter direction:
\begin{equation}
    e^{i\varphi} := \frac{1}{|\bm{\psi}_1|}
    \bm{\psi}_1^\dag \bm{x}^{(1)}
    ~~ ~~\text{and} ~~ ~~e^{i\varphi_0} := \frac{1}{|\bm{\psi}_1|} \bm{\psi}_1^\dag\bm{x}_0^{(1)}
\end{equation} in the thermodynamic limit. Using the generalized Watanabe-Strogatz transformation in Eq.~(\ref{eq:Tanaka_WS}), we obtain
\begin{flalign}
e^{i\varphi} &:= \frac{1}{|\bm{\psi}_1|}
    \bm{\psi}_1^\dag \bm{x}^{(1)} = \frac{1}{|\bm{\psi}_1|} \frac{\bm{\psi}_1^\dag \bm{A}\bm{x}^{(1)}_0 + |\bm{\psi}_1|^2 }{|\bm{\psi}_1|e^{i\varphi_0}+1} \notag \\
    &= \frac{1}{|\bm{\psi}_1|}\frac{\bm{\psi}_1^\dag\bm{x}^{(1)}_0+ |\bm{\psi}_1|^2 }{|\bm{\psi}_1|e^{i\varphi_0}+1} =\frac{e^{i\varphi_0}+ |\bm{\psi}_1| }{|\bm{\psi}_1|e^{i\varphi_0}+1}
    \label{eq:dist1}
\end{flalign} with $\bm{A}=\bm{H}^{1/2}\bm{U}$ and $\bm{H}^{1/2}\bm{\psi}_1 = \bm{\psi}_1$ where we can set $\bm{U}=I_M$ as long as $\bm{x}^{(1)}_0$ are uniformly distributed on $\mathbb{S}^{M}$. Algebraically rearranging Eq.~(\ref{eq:dist1}) and following the same argument in Sec.~\ref{Sec:review}, we obtain
\begin{equation}
    e^{i\varphi_0} = \frac{e^{i\varphi}-|\bm{\psi}_1|}{1-|\bm{\psi}_1|e^{i\varphi}}.
\end{equation} Denoting $\varphi = T(\varphi_0)$ and $\varphi_0 = T^{-1}(\varphi)$ (inverse transformation), one obtains $d(T_*\mu)(\varphi) = f(\varphi)d\varphi$ where $f(\varphi):=\frac{1}{2\pi} \partial_\varphi T^{-1}(\varphi)$ is the phase distribution function, and $d\mu(\varphi_0) = \frac{1}{2\pi}d\varphi_0$ since $\bm{x}_0^{(1)}$ are uniformly distributed on $\mathbb{S}^{M}$. Then, we obtain
\begin{flalign}
    f(\varphi)&=\frac{1}{2\pi}\partial_\varphi T^{-1}(\varphi) = \frac{1}{2\pi i} \bigg( \frac{i e^{i\varphi}}{e^{i\varphi}-|\bm{\psi}_1|} +\frac{i |\bm{\psi}_1|e^{i\varphi}}{1-|\bm{\psi}_1|e^{i\varphi}} \bigg) \notag \\
    &=\frac{1}{2\pi}\frac{1-|\bm{\psi}_1|^2}{1-2 |\bm{\psi}_1|\cos\varphi+|\bm{\psi}_1|^2} \label{eq:normalized_Poisson}
\end{flalign} which is the normalized Poisson kernel distribution. This reminds us of the Ott-Antonsen manifold for the KS oscillators on the unit circle of $\mathbb{C}^1$ where oscillators' phases are distributed according to the normalized Poisson kernel in Eq.~(\ref{eq:Poisson_kernel}). For the higher dimensional case $\mathbb{C}^2$, we also find that the phase distribution along the generalized WS variable satisfies the normalized Poisson kernel. In Fig.~\ref{Fig:micro_complex} (f), the histogram of the distribution (blue bars) of $\{ \varphi_j := \frac{1}{|\bm{\psi}_1|} \bm{\psi}_1^\dag\bm{x}_j^{(1)} \}_{j=1}^{N}$ is shown for $N=500$ and $A=0.35$. The numerical results for the finite-sized ensemble fits well to the analytical prediction (red curve) in Eq.~(\ref{eq:normalized_Poisson}).

From the above results, it is anticipated that the oscillators on $\mathbb{S}^M$ are distributed according to the higher-dimensional Poisson kernel in $\mathbb{C}^M$. Assume that the oscillators on $\mathbb{S}^M$ are distributed by
\begin{flalign}
 f_\text{complex}(\bm{x},\bm{\psi};t) = \frac{1}{S_{2M}}\frac{1-|\bm{\psi}|^2}{|\bm{\psi}-\bm{x}|^{2M}} \label{eq:higher_poisson}
\end{flalign} for $\bm{\psi} \in \mathbb{C}^M$ and for $\bm{x} \in \mathbb{S}^M$. Then, we obtain for the Kuramoto order parameter
\begin{flalign}
 \bm{m}(t) &= \int_{|\bm{x}|=1} \bm{x} f(\bm{x},\bm{\psi}) d\bm{x} \notag \\
 &= \frac{1-|\bm{\psi}|^2}{S_{2M}} \int_{|\bm{x}|=1} \frac{\bm{x}}{(1+|\bm{\psi}|^2-2 \text{Re}\braket{\bm{\psi}}{\bm{x}})^M} d\bm{x} \notag \\
 & = \frac{1-|\bm{\psi}|^2}{S_{2M}} \int_{|\eta| \leq 1} \int_{|\bm{n}|=\sqrt{1-|\eta|^2}} \frac{\eta \hat{\bm{\psi}} + \bm{n} }{(1+|\bm{\psi}|^2-2 \text{Re}\braket{\bm{\psi}}{\bm{x}})^M} d\bm{n} \frac{1}{\sqrt{1-|\eta|^2}}d\eta
\end{flalign} where $\hat{\bm{\psi}} := \frac{\bm{\psi}}{|\bm{\psi}|} $ is a unit vector and $\braket{\bm{n}}{\bm{\psi}}=0$. Here, we decompose the unit vector on the sphere into $\bm{x} = \eta \hat{\bm{\psi}}+\bm{n}$~\cite{Tanaka_2014}. Then, the integral can be written as
\begin{flalign}
 \bm{m}(t) &= \frac{1-|\bm{\psi}|^2}{S_{2M}} \hat{\bm{\psi}}  \int_{|\eta| \leq 1} \int_{|\bm{n}|=\sqrt{1-|\eta|^2}} \frac{\eta}{(1+|\bm{\psi}|^2-2 |\bm{\psi}| \text{Re}\eta)^M} d\bm{n} \frac{1}{\sqrt{1-|\eta|^2}} d\eta \notag \\
 &= \frac{ S_{2M-2}}{S_{2M}}(1-|\bm{\psi}|^2) \hat{\bm{\psi}}  \int_{|\eta| \leq 1} \frac{\eta}{(1+|\bm{\psi}|^2-2 |\bm{\psi}| \text{Re}\eta)^M} d\bm{n} \frac{(1-|\eta|^2)^{\frac{2M-3}{2}}}{\sqrt{1-|\eta|^2}} d\eta \notag \\
 &= \frac{ S_{2M-2}}{S_{2M}}(1-|\bm{\psi}|^2) \hat{\bm{\psi}} \int_0^1 \int_0^{2\pi} \frac{r e^{i\theta}}{(1+|\bm{\psi}|^2-2|\bm{\psi}|r \cos\theta)^M} (1-r^2)^{M-2} rd\theta dr \notag \\
 &= \bm{\psi}(t)
 \end{flalign} which is consistent with the result above, i.e., $\bm{m}(t) = \bm{\psi}(t)$ for the complex space. Thus, we can expect that the oscillators for the complex spaces are distributed according to Eq.~(\ref{eq:higher_poisson}), i.e., the higher-dimensional normalized Poisson kernel. However, for the real spaces, $\mathbb{R}^M$, the normalized Poisson kernel is given by
\begin{equation}
f_\text{real}(\bm{x} ,\bm{\psi};t) = \frac{1}{S_M} \frac{1-|\bm{\psi}|^2}{|\bm{\psi}-\bm{x}|^M} \label{eq:higher_poisson_real}
\end{equation} such that the Kuramoto order parameter reads
\begin{flalign}
\bm{m}(t) &= \frac{1-|\bm{\psi}|^2}{S_M} \int_{|\bm{x}|=1} \frac{\bm{x}}{(1+|\bm{\psi}|^2-2 \braket{\bm{\psi}}{\bm{x}})^{M/2}} d\bm{x} \notag \\
&= \frac{1-|\bm{\psi}|^2}{S_M} \int_{-1}^{1}\int_{|\bm{n}|=\sqrt{1-\eta^2}} \frac{\eta \hat{\bm{\psi}}  + \bm{n} }{(1+|\bm{\psi}|^2-2 \braket{\bm{\psi}}{\bm{x}})^{M/2}} d \bm{n} \frac{d\eta}{\sqrt{1-\eta^2}} \notag \\
&= \frac{S_{M-1}}{S_M}(1-|\bm{\psi}|^2) \hat{\bm{\psi}} \int_{-1}^{1} \frac{\eta}{(1+|\bm{\psi}|^2-2 |\bm{\psi}| \eta)^{M/2}} \frac{(1-\eta^2)^{\frac{M-2}{2}}}{\sqrt{1-\eta^2}}d\eta \notag \\
&=\bm{\psi}(t)
\end{flalign} which is inconsistent with Eq.~(\ref{eq:order_real}). This implies that the distribution of the oscillators on $\mathbb{S}^M$ for $\mathbb{R}^M$ is not given by Eq.~(\ref{eq:higher_poisson_real}) with the given  $\bm{\psi}$, i.e., the higher-dimensional normalized Poisson kernel. It was reported that the real oscillators distributed according to the higher-dimensional Poisson kernel with a given WS variable $\bm{\psi}$ in the thermodynamic limit satisfy the OA equations introduced in Refs.~\cite{g_KM3,g_KM4}. Therein, the spherical harmonics expansion was exploited for the oscillator distribution function, as Ott and Antonsen used Fourier expansion for 2D real space~\cite{OA1,OA2}. For example, in Ref.~\cite{g_KM3}, the distribution of 3D real oscillators was assumed to be
\begin{flalign}
    f_\text{real}(\theta,\phi;t) &= \sum_{l=0}^{\infty}\sum_{m=-l}^{l}f_{lm}Y_{lm}(\theta,\phi) \notag \\
    &= \frac{1}{S_3} \bigg( 1 + 4\pi \sum_{l=0}^{\infty}\sum_{m=-l}^{l} \rho^l \overline{Y_{lm}}(\Theta,\Phi) Y_{lm}(\theta,\phi)  \bigg)
\end{flalign} where $f_{lm} = \rho^l \overline{Y_{lm}}(\Theta,\Phi)$ is taken as a generalied OA ansatz and $Y_{lm}(\theta,\phi)$ are spherical harmonics~\cite{hassani2013mathematical}. To obtain the phase distribution function, one can use
\begin{flalign}
    \frac{4\pi}{2l+1}\sum_{m=-l}^{l}Y_{lm}(\hat{y})\overline{Y_{lm}}(\hat{x}) = P_l(\hat{x}\cdot \hat{y}) ~~ ~~ \text{and} ~~ ~~ \sum_{l=0}^{\infty} y^l P_l(x) = \frac{1}{\sqrt{1+y^2-2xy}}
\end{flalign} where $P_l(x)$ are Legendre polynomials. Using the above relations, we can reach
\begin{flalign}
     f_\text{real}(\theta,\phi;t) &= \frac{1}{S_3} \bigg(1 + 4\pi \sum_{l=1}^{\infty} \frac{2l+1}{4\pi} \rho^l P_l(\hat{r}\cdot \hat{\rho})  \bigg) =\frac{1}{S_3} \bigg( 2\sum_{l=0}^{\infty} l \rho^l  P_l(\hat{r}\cdot \hat{\rho}) + \sum_{l=0}^{\infty}\rho^l P_l(\hat{r}\cdot \hat{\rho})  \bigg) \notag \\
     &=\frac{1}{S_3} \bigg(  2\rho\frac{\partial}{\partial\rho}\sum_{l=0}^{\infty} \rho^l  P_l(\hat{r}\cdot \hat{\rho}) + \sum_{l=0}^{\infty}\rho^l P_l(\hat{r}\cdot \hat{\rho}) \bigg) \notag \\
     &= \frac{1}{S_3} \bigg( 2\rho \frac{\partial}{\partial\rho} \frac{1}{\sqrt{1+\rho^2-2\hat{r}\cdot \hat{\rho}}} +\frac{1}{\sqrt{1+\rho^2-2\hat{r}\cdot \hat{\rho}}} \bigg)
\end{flalign} where $\bm{\rho} = \rho (\sin\Theta\cos\Phi,\sin\Theta\sin\Phi,\cos\Theta)^\top \in \mathbb{R}^3$ is the OA variable (in our notation, $\bm{\psi}(t)=\bm{\rho}(t)$) and the microscopic oscillator is represented as $\bm{r} =  (\sin\theta\cos\phi,\sin\theta\sin\phi,\cos\theta)^\top \in \mathbb{S}^3$ in the thermodynamic limit. Finally, the oscillator distribution is given as~\cite{g_KM3}
\begin{flalign}
    f_\text{real}(\theta,\phi,t) = \frac{1}{4\pi}\frac{1-\rho^2}{(1+\rho^2-2\rho \hat{\rho} \cdot \hat{r})^{3/2}} = \frac{1}{S_3} \frac{1-|\bm{\rho}|^2}{|\bm{\rho}-\bm{r}|^3}
\end{flalign} as in Eq.~(\ref{eq:higher_poisson_real}). However, it was reported in Ref.~\cite{g_KM3,g_KM4} that the OA variable is then governed by
\begin{flalign}
     \dot{\bm{\rho}} = \bm{\Omega}  \bm{\rho} +\frac{1}{2}(1+|\bm{\rho}|^2)(\bm{K}\bm{\Gamma}) - \big[ \bm{\rho}^\top (\bm{K}\bm{\Gamma}) \big] \bm{\rho}
\end{flalign} which looks different from Eq.~(\ref{eq:WS_dynamics_Tanaka}). For the higher dimensional real space, see the OA equation in Ref.~\cite{g_KM4}. In conclusion, both for the higher-dimensional complex and real spaces, the oscillators are distributed according to the higher-dimensional Poisson kernel in the Ott-Antonsen manifold. For the complex oscillators, the model and the generalized WS transformation in Ref.~\cite{Tanaka_2014} give the correct way, and the Kuramoto order parameter exactly coincides with the OA variable. However, to obtain the Poisson kernel distribution for the real spaces in the OA manifold, we need to consider the spherical harmonics expansion and the governing equations reported in Refs.~\cite{g_KM3,g_KM4} for the real spaces, and in that manifold, the Kuramoto order parameter exactly equals to the OA variable. Nevertheless, we will exploit the model in Ref.~\cite{Tanaka_2014} for the higher dimensional real spaces below.

\section{\label{sec:real_system}Chimera Dynamics in Two-population Networks for $\mathbb{R}^4$}

In this section, as a comparison, we investigate the system of identical generalized Kuramoto-Sakaguchi oscillators in two-population networks for $\mathbb{K}=\mathbb{R}$ in the thermodynamic limit. The WS variables $\bm{\psi}_a(t) \in \mathbb{R}^M$ are governed by
\begin{flalign}
\dot{\bm{\psi}}_a &= - \bm{\psi}_a \bm{g}_a^\dag \bm{\psi}_a + \bm{g}_a \notag \\
&= - \bm{\psi}_a \bm{g}_a^\top \bm{\psi}_a + \bm{g}_a \label{eq:WS_two_population_real}
\end{flalign} for $a=1,2$ where $\dag = \top$ (i.e., hermitian adjoint=transpose) for $\mathbb{K}=\mathbb{R}$. In this system, the mean-field forcing should be different since the Kuramoto order parameter not exactly coincides with the WS variable as in Eq.~(\ref{eq:order_real}), i.e., $\bm{m}_a(t) = h(|\bm{\psi}_a|^2,M)\bm{\psi}_a(t)$. Thus, we have to consider
\begin{equation}
    \bm{g}_a := \mu \bm{K}h(|\bm{\psi}_a|^2,M)\bm{\psi}_a(t) +\nu \bm{K} h(|\bm{\psi}_b|^2,M)\bm{\psi}_b(t) 
\end{equation} for $(a,b)=(1,2)$ or $(2,1)$. Here, the coupling matrix $\bm{K}$ is a suitable rotational matrix introduced in Sec.~\ref{sec:BF}. In this section, we use following notations: $\bm{\psi}_1 = (x_1,...,x_M)^\top$ and $\bm{\psi}_2 = (y_1,...,y_M)^\top$ where $x_i, y_i \in \mathbb{R}$ for $i=1,...,M$.

To get an overview of the observable chimera states, we first investigate the reduced dynamics, i.e., the dynamics of the magnitude of $\bm{\psi}_a$. Considering Eq.~(\ref{eq:WS_two_population_real}), we obtain
\begin{flalign}
\dot{\rho}_1 &=\frac{1-\rho_1^2}{\rho_1}\big(\mu h(\rho_1^2,M)\rho_1^2\cos\alpha +\nu h(\rho_2^2,M) \xi \big) \notag \\
\dot{\rho}_2 &= \frac{1-\rho_2^2}{\rho_2}\bigg(\mu h(\rho_2^2,M)\rho_2^2\cos\alpha + \nu h(\rho_1^2,M)\big( \xi\cos2\alpha +\sin2\alpha \sqrt{\rho_1^2\rho_2^2-\xi^2} \big)  \bigg) \notag \\
\dot{\xi} &=\mu h(\rho_1^2,M) \bigg( \xi(1-\rho_1^2)\cos\alpha +\sin\alpha \sqrt{\rho_1^2\rho_2^2-\xi^2} \bigg) \notag \\
&~~ ~~+ \mu h(\rho_2^2,M) \bigg( \xi(1-\rho_2^2)\cos\alpha -\sin\alpha \sqrt{\rho_1^2\rho_2^2-\xi^2} \bigg) \notag \\
& ~~ ~~+ \nu h(\rho_1^2,M)\bigg((\rho_1^2-\xi^2)\cos2\alpha -\sin2\alpha\sqrt{\rho_1^2\rho_2^2-\xi^2} \bigg)+\nu h(\rho_2^2,M)(\rho_2^2-\xi^2) \label{eq:magnitude_dynamics_real}
\end{flalign} where $\rho_a := |\bm{\psi}_a|$ for $a=1,2$ and $\xi := \braketmatrix{\bm{\psi}_1}{\bm{K}}{\bm{\psi}_2}$ indicates the cross term. For the derivation of Eq.~(\ref{eq:magnitude_dynamics_real}), see \ref{append:derivation2}. Note that the dynamics of the magnitude of $\bm{\psi}_a$ here depends on the dimension $M$ whereas it is independent of $M$ for the complex space as in Eq.~(\ref{eq:OA_two_pop}).

\begin{figure}[t!]
\centering
\includegraphics[width=0.65\linewidth]{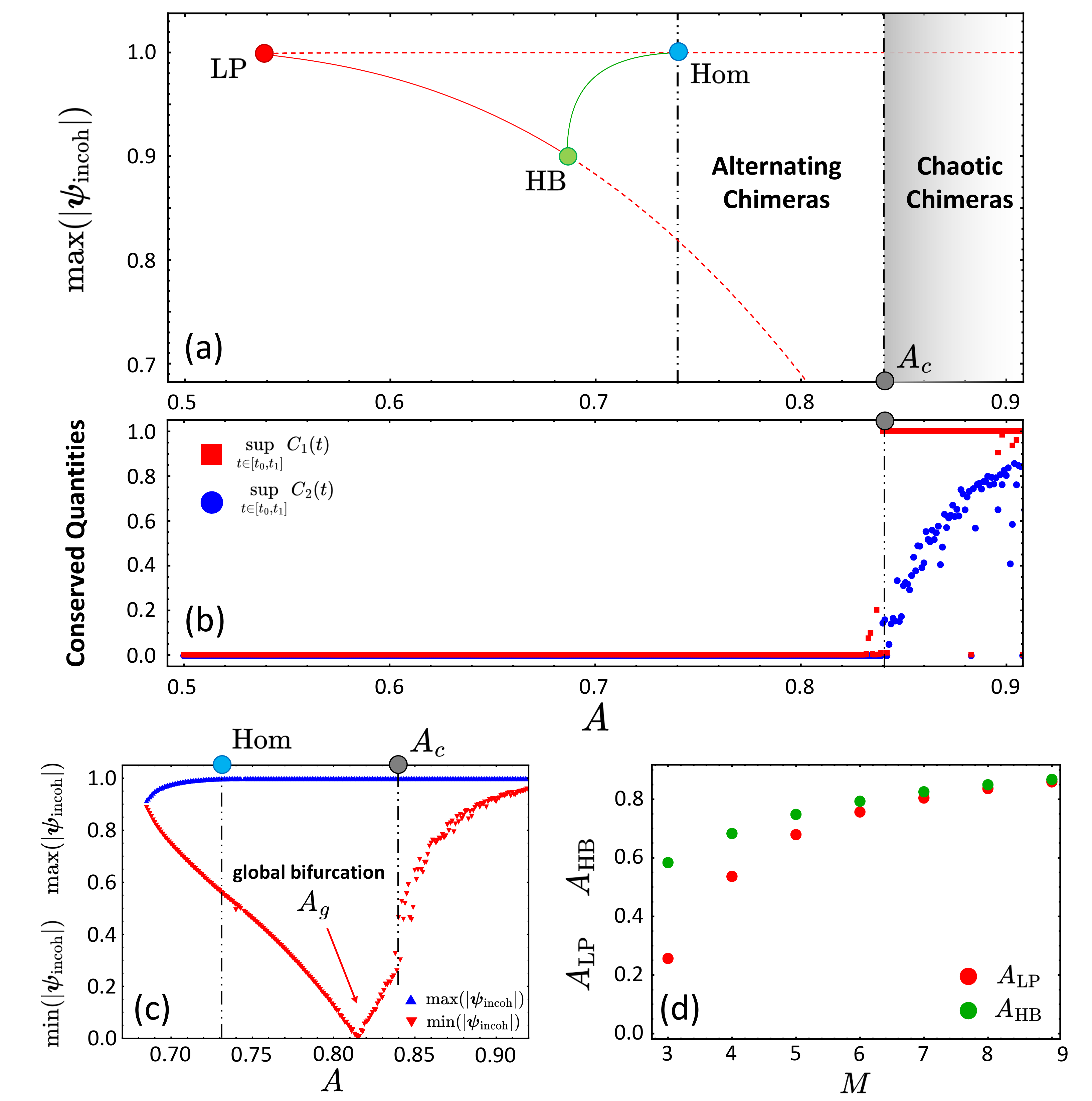}
\caption{(a) Bifurcation diagram of stable (solid curve) and unstable (dashed curve) chimera states in two-population networks. Red: stationary chimeras and Green: breathing chimera states. LP: limiting point bifurcation, HB: supercritical Hopf bifurcation, Hom: homoclinic bifurcation, and $A_c$ from which on a chaotic chimera appears. (b) Conserved quantities as a function of the parameter $A$. $t_0 = 20,000$ and $t_1 =10^5$. (c) Maximum and minimum values of the magnitude of the WS variable $\bm{\psi}_1$ as a function of the parameter $A$. All the results are obtained for $\mathbb{R}^4$. (d) Bifurcation points $A_\text{LP}$ and $A_\text{HB}$ as a function of the dimension $M$.} 
\label{Fig:bif_real}
\end{figure}

Below, we focus on the dynamics for $\mathbb{R}^4$ unless stated otherwise. In Fig.~\ref{Fig:bif_real} (a), a bifurcation diagram from Eq.~(\ref{eq:magnitude_dynamics_real}) is depicted as the parameter $A$ varies by assuming again $\rho_2=1$. As similar to Fig.~\ref{Fig:symmetry_broken} (a), a stable (red, solid) and unstable (red, dashed) stationary chimeras emerge in a limiting point (LP) bifurcation, however, at rather larger value of $A_\text{LP}$ compared to that of $\mathbb{C}^2$. Then, the stable stationary chimera state undergoes a supercritical Hopf bifurcation, producing a stable limit-cycle solution, i.e., a breathing chimera state (green, solid). The breathing chimera undergoes a homoclinic bifurcation as its period increases indefinitely. The dynamics of the magnitude of the WS variable in Eq.~(\ref{eq:magnitude_dynamics_real}) only shows this scenario. However, we also observe the periodic alternating chimera state in the full component-dynamics from Eq.~(\ref{eq:WS_two_population_real}). For a given $A \in (A_\text{Hom},A_c)$, a periodic alternating chimera state can be observed, apparently following  $\bm{\psi}_1(t) = \bm{\psi}_2(t-\frac{T}{2})$ and $x_i(t) = y_i(t-\frac{T}{2})$ for $i=1,...,4$ where $T$ is the period of $\bm{\psi}_a$ for $a=1,2$. Likewise, we numerically find that there are two conserved quantities along the trajectory of the stationary, breathing and periodically alternating chimera states. The conserved quantities are similar to those in the complex space. First, we define angular variables $\theta_1 := \tan^{-1}\frac{x_2}{x_1}$, $\theta_2 := \tan^{-1}\frac{x_4}{x_3}$, $\phi_1 := \tan^{-1}\frac{y_2}{y_1}$, $\phi_2 := \tan^{-1}\frac{x_4}{x_3}$. Then, Equation (\ref{eq:WS_two_population_real}) is invariant under the rotational transformation such as
\begin{equation}
    Q := \begin{pmatrix}
 R(-\theta_1) & \bm{0}  \\
\bm{0} & R(-\theta_2)
\end{pmatrix} ~~ ~~ \text{where} ~~ ~~ R(\theta) = \begin{pmatrix}
\cos\theta & \sin\theta  \\
-\sin\theta & \cos\theta 
\end{pmatrix} \notag
\end{equation} where $Q \in \mathbb{R}^{4 \times 4}$, $R(\theta) \in \mathbb{R}^{2 \times 2}$, and $\bm{0} \in \mathbb{R}^{2 \times 2}$ is the zero-matrix. Therefore, we can define the phase difference $\Theta_1 :=  \phi_1-\theta_1$ and $\Theta_2 :=  \phi_2- \theta_2$. The first conserved quantity is
\begin{flalign}
C_1:= \sin\big(\Theta_1-\Theta_2 \big) \label{eq:con1_real}
\end{flalign} and one can obtain $C_1(t)=0$ for all $t$ for $A<A_c$ along a chimera trajectory. We also find numerically another conserved quantity that reads
\begin{flalign}
C_2 &:=  \sum_{k=1}^4 x_k^2  \sum_{k=1}^4 y_k^2- \bigg( \sqrt{x_1^2+x_2^2}\sqrt{y_1^2+y_2^2}+ \sqrt{x_3^2+x_4^2} \sqrt{y_3^2+y_4^2} \bigg)^2 \label{eq:con2_real}
\end{flalign} which arises from the relation
\begin{flalign}
    \braket{\bm{\psi}_1}{\bm{\psi}_2} &= |\bm{\psi}_1| |\bm{\psi}_2|\cos\Theta \notag \\
    &= \sqrt{x_1^2+x_2^2}\sqrt{y_1^2+y_2^2}\cos\Theta   + \sqrt{x_3^2+x_4^2} \sqrt{y_3^2+y_4^2}\cos\Theta 
\end{flalign} where $\Theta:=\Theta_1=\Theta_2$ since $C_1=0$. In Fig.~\ref{Fig:bif_real} (b), $\sup_{t \in [t_0,t_1]}C_1(t)$ and $\sup_{t \in [t_0,t_1]}C_1(t)$ are shown as the parameter $A$ varies for $t_0=2 \times 10^4$ and $t_1=10^5$. This numerically verifies that the two quantities are indeed conserved along a chimera trajectory for $A<A_c$. On the other hand, we observe the chimera state that breaks the two conserved quantities for $A>A_c$ (Fig.~\ref{Fig:bif_real} (b)). This chimera state, similar as for $\mathbb{C}^2$ (cf.  Fig.~\ref{Fig:chaotic_detail}), also shows aperiodic motion of the components of the WS variables $\bm{\psi}_a$ for $a=1,2$ (not shown here). Hence, for $\mathbb{R}^4$, we observe as well that aperiodic chimera dynamics appears via breaking the conserved quantities in Eqs.~(\ref{eq:con1_real}-\ref{eq:con2_real}).

Furthermore, also in the real space, a global bifurcation appears involving the completely incoherent state. However, this global bifurcation is observed for the periodic alternating chimera state for $A>A_\text{Hom}$ rather than the breathing chimeras, which is somewhat different from the case of the complex space (Fig.~\ref{Fig:symmetry_broken}). In Fig.~\ref{Fig:bif_real} (c), the maximum and the minimum values of $|\bm{\psi}_\text{incoh}|$ are depicted with the parameter $A$ changed. As $A$ increases, for breathing chimera state, $\text{min}|\bm{\psi}_\text{incoh}|$ continuously decreases, and then seamlessly connects to that of the periodic alternating chimera state for $A>A_\text{Hom}$. The phase variables $\Theta(t)$ of these chimeras evolve within some interval smaller than $\mathbb{T}$, similar to Fig.~\ref{Fig:symmetry_broken} (d-1). Further increasing $A$, the alternating chimera state touches the completely incoherent state of $|\bm{\psi}_1|=0$ and its phase variable monotonically increases as a function of time as in Fig.~\ref{Fig:symmetry_broken} (d-2).

Finally, we note that the dynamics of the magnitude of the WS variables depends on the dimension $M$. For small $A$, the scenarios of the emergence of the chimera states in low-dimensional real spaces reveals that the stationary chimera state is born/annihilated at a limiting point bifurcation ($A_\text{LP}$) and it undergoes a supercritical Hopf bifurcation at $A=A_\text{HB}$. However, Figure~\ref{Fig:bif_real} (d) displays the parameter interval between the limiting point and the Hopf bifurcation, i.e., $(A_\text{LP},A_\text{HB})$. It decreases as $M$ increases. This indicates that for the higher-dimensional real systems, it becomes harder to obtain stationary chimera states. Not only for the stationary chimeras, but other chimera states also are hardly observed within our numerical ability for higher dimensional spaces.

\section{\label{sec:Conlcusion}Summary and Outlook}


In this paper, we explored a system of identical generalized Kuramoto-Sakaguchi oscillators defined on the surface of the unit sphere in two-population networks. We exploited the model proposed in Ref.~\cite{Tanaka_2014} to take advantage of the extended Watanabe-Strogatz transformation that describes the macroscopic dynamics of the system. First, we introduced a suitable coupling matrix both for the real and complex spaces, in line with the standard KS model defined on the unit circle and its phase-lag parameter, which becomes relevant to determine the Benjamin-Feir instability point. For the 2D complex space $\mathbb{C}^2$, particularly in the thermodynamic limit, stationary chimeras are created/destroyed at a limiting point bifurcation, and the stable chimera undergoes a supercritical Hopf bifurcation that produces a stable breathing chimera state. In this system, the breathing chimera states undergo a global bifurcation involving the completely incoherent state, which changes their dynamics in terms of the phase variables. Beyond this global bifurcation, the periodic alternating chimera dynamics appears seamlessly connecting to the breathing chimeras in terms of the minimum value of the order parameter magnitude. These three types of chimera states possess two conserved quantities along their trajectory in phase space. However, when the coupling strength is further weakened, the chimera trajectory shows componentwise aperiodic dynamics via the breaking of the conserved quantities. 

To back up this further, we also explore a finite-sized ensemble of identical generalized KS oscillators in two-populations for $\mathbb{C}^2$. The finite-sized system exhibits states corresponding to each chimera state mentioned above in terms of the Kuramoto order parameter. Furthermore, the microscopic dynamics conserves the quantities defined by the components of the Kuramoto order parameters, which then break for larger parameter values of $A$. Finally, we compared this result of $\mathbb{C}^2$ to the dynamics for $\mathbb{R}^4$ in the thermodynamic limit. The chimera states in the real space again follow a similar scenario, including stationary and breathing chimeras, conserved quantities, and also the breaking of the latter to induce componentwise aperiodic chimera dynamics. However, in this case, the global bifurcation occurs for the alternating chimera states rather than for the breathing chimera states.

A system of coupled oscillators in two-population networks has been considered by many researchers for the study of chimera states, as well as many variations of this topology. Accordingly, one might explore chimera states of generalized Kuramoto-Sakaguchi oscillators in three- and multi-population networks. To consider more realistic situations, heterogeneities on the system could be imposed on the oscillator ensemble. It would be also an interesting application to study chimeras of heterogeneous generalized KS oscillators. For the heterogeneous natural frequency distribution, one should consider the macroscopic dynamics in Refs.~\cite{Tanaka_2014,g_KM4} for complex and real spaces, respectively. Furthermore, it is also possible to arrange the generalized Kuramoto oscillators along a ring geometry with nonlocal coupling, and then investigate chimera states~\cite{short_chimera}. Apart from these, there could be many other applications of the generalized Kuramoto-Sakaguchi model that provides us with further understandings of chimera dynamics.

\appendix

\section{\label{append:order_parameter}Derivation of Eqs.~(\ref{eq:order_real}-\ref{eq:order_complex})}

In Ref.~\cite{Tanaka_2014}, the author obtained Eqs.~(\ref{eq:order_real}-\ref{eq:order_complex}) by directly calculating the definition of the order parameter in the thermodynamic limit. In this Appendix, we adopt a different approach to obtain the same result, following a similar method as introduced by Pikovsky and Rosenblum.~\cite{pikovsky_WS1,pikovsky_WS2}. Consider the order parameter defined in Eq.~(\ref{eq:COM}) with the WS transformation in Eq.~(\ref{eq:Tanaka_WS}):
\begin{flalign}
\bm{m}(t) &= \frac{1}{N}\sum_{k=1}^{N}\bm{x}_k(t) = \frac{1}{N}\sum_{k=1}^{N}\frac{\bm{A}\bm{x}_{0,k} + \bm{b}}{\bm{b}^\dag \bm{A}\bm{x}_{0,k}+1}. \notag 
\end{flalign} Substituting $\bm{A}=\bm{H}^{1/2}\bm{U}$ and $\bm{H}^{1/2}\bm{b}=\bm{b}$ into the above equation, we obtain 
\begin{flalign}
\bm{b}^\dag \bm{m} &= \frac{1}{N}\sum_{k=1}^{N} (\bm{b}^\dag \bm{H}^{1/2}\bm{U} \bm{x}_{0,k} + |\bm{b}|^2) \frac{1}{1+ \bm{b}^\dag \bm{U} \bm{x}_{0,k}} \notag \\
&=\frac{1}{N}\sum_{k=1}^{N} (\bm{b}^\dag \bm{U} \bm{x}_{0,k} + |\bm{b}|^2) \sum_{ \ell =0}^{\infty}(-1)^{\ell} (\bm{b}^\dag \bm{U}\bm{x}_{0,k})^{\ell} \notag \\
&= -\frac{1}{N}\sum_{k=1}^N \sum_{\ell=0}^\infty (-1)^{\ell+1}(\bm{b}^\dag \bm{U}\bm{x}_{0,k})^{\ell+1} + |\bm{b}|^2\frac{1}{N}\sum_{k=1}^N \sum_{\ell=0}^\infty (-1)^{\ell}(\bm{b}^\dag \bm{U}\bm{x}_{0,k})^{\ell}
\notag \\
&=|\bm{b}|^2 \bigg( 1 + (1-|\bm{b}|^{-2}) \sum_{\ell=2}^{\infty}(-1)^{\ell} C_\ell \bigg) \label{eq_append:b_m}
\end{flalign} where $C_\ell := \frac{1}{N}\sum_{k=1}^{N}(\bm{b}^\dag \bm{U} \bm{x}_{0,k} )^\ell$ and $C_1=0$. Taking the thermodynamic limit with uniformly distributed constants of motion, we can set $\bm{U}=I_M$ and write $C_\ell$ as
\begin{flalign}
C_\ell = \begin{dcases}
   \frac{1}{S_M}\int_{|\bm{x}_0|=1} (\bm{b}^\dag \bm{x}_0)^\ell d\bm{x}_0, & \text{for}~~\mathbb{K}=\mathbb{R} \\ \\
     \frac{1}{S_{2M}}\int_{|\bm{x}_0|=1}(\bm{b}^\dag \bm{x}_0)^\ell d\bm{x}_0, & \text{for}~~\mathbb{K}=\mathbb{C}
  \end{dcases} \label{eq_append:C-l}
\end{flalign} Let us first consider the real space, i.e., $\mathbb{K}=\mathbb{R}$. Using the spherical coordinate systems, one can write a position of an oscillator on the surface of the unit ball in $\mathbb{R}^M$ as $x_1 = \sin\theta_1\sin\theta_2 \cdots \sin\theta_{M-2}\cos\phi$, $x_2 = \sin\theta_1\sin\theta_2 \cdots \sin\theta_{M-2}\sin\phi$, ..., $x_{M-2} = \sin\theta_1\sin\theta_2\cos\theta_3$, $x_{M-1} = \sin\theta_1\cos\theta_2$, and $x_{M} = \cos\theta_1$ where $\theta_1,...,\theta_{M-2} \in [0,\pi]$ and $\phi \in [0,2\pi]$. Here, $x_M$ corresponds to the $z$-axis, for example, in the 3D space. We call this axis the $M$-axis and we refer to the plane perpendicular to the $M$-axis as the $M^{\perp}$-(hyper)plane throughout this paper. Without loss of generality, we can assume that $\bm{b}$ is aligned along the $M$-axis, i.e., $\bm{b} = |\bm{b}| \hat{x}_M$. Consequently, we obtain
\begin{flalign}
C_\ell &= \frac{1}{S_M}|\bm{b}|^\ell \int_{|\bm{x}_0|=1} \cos^{\ell}\theta_1 d^MV =\frac{|\bm{b}|^\ell}{S_M} S_{M-1} \int \cos^\ell\theta_1\sin^{M-2}\theta_1 d\theta_1 \notag \\
&=|\bm{b}|^\ell \frac{1+(-1)^\ell}{2\sqrt{\pi}} \frac{\Gamma\big(\frac{\ell+1}{2}\big) \Gamma\big(\frac{M}{2}\big)}{\Gamma\big(\frac{\ell+M}{2}\big)} \notag
\end{flalign} which leads to
\begin{flalign}
\sum_{\ell=2}^\infty (-1)^\ell C_\ell &= |\bm{b}|^2 \frac{\Gamma\big(\frac{M}{2}\big)}{2 \Gamma\big(1+\frac{M}{2}\big)}   \, _2F_1(1,\frac{3}{2};\frac{M+2}{2};|\bm{b}|^2). \notag
\end{flalign} Finally, Equation (\ref{eq_append:b_m}) can be written as
\begin{flalign}
\bm{b}^\dag \bm{m} &= |\bm{b}|^2 \bigg(1+ \frac{|\bm{b}|^2-1}{M} \, _2F_1(1,\frac{3}{2};\frac{M+2}{2};|\bm{b}|^2)  \bigg) \notag \\
&= \bm{b}^\dag \bm{b} \frac{M-1}{M} \, _2F_1(\frac{1}{2},1;\frac{M+2}{2};|\bm{b}|^2). \notag
\end{flalign} Hence, we can assume that
\begin{flalign}
\bm{m}(t) = \frac{M-1}{M} \, _2F_1(\frac{1}{2},1;\frac{M+2}{2};|\bm{b}|^2) \bm{b}(t).
\end{flalign} For $\mathbb{K} = \mathbb{C}$, obtaining $\bm{m}(t) = \bm{b}(t)$ is straightforward as the coefficients $C_\ell = 0$ for $\ell \neq 0$.

\section{\label{append:3d_Benjamin} Benjamin-Feir Instability Point for $\mathbb{R}^3$}

\begin{figure}[t!]
\centering
\includegraphics[width=0.7\linewidth]{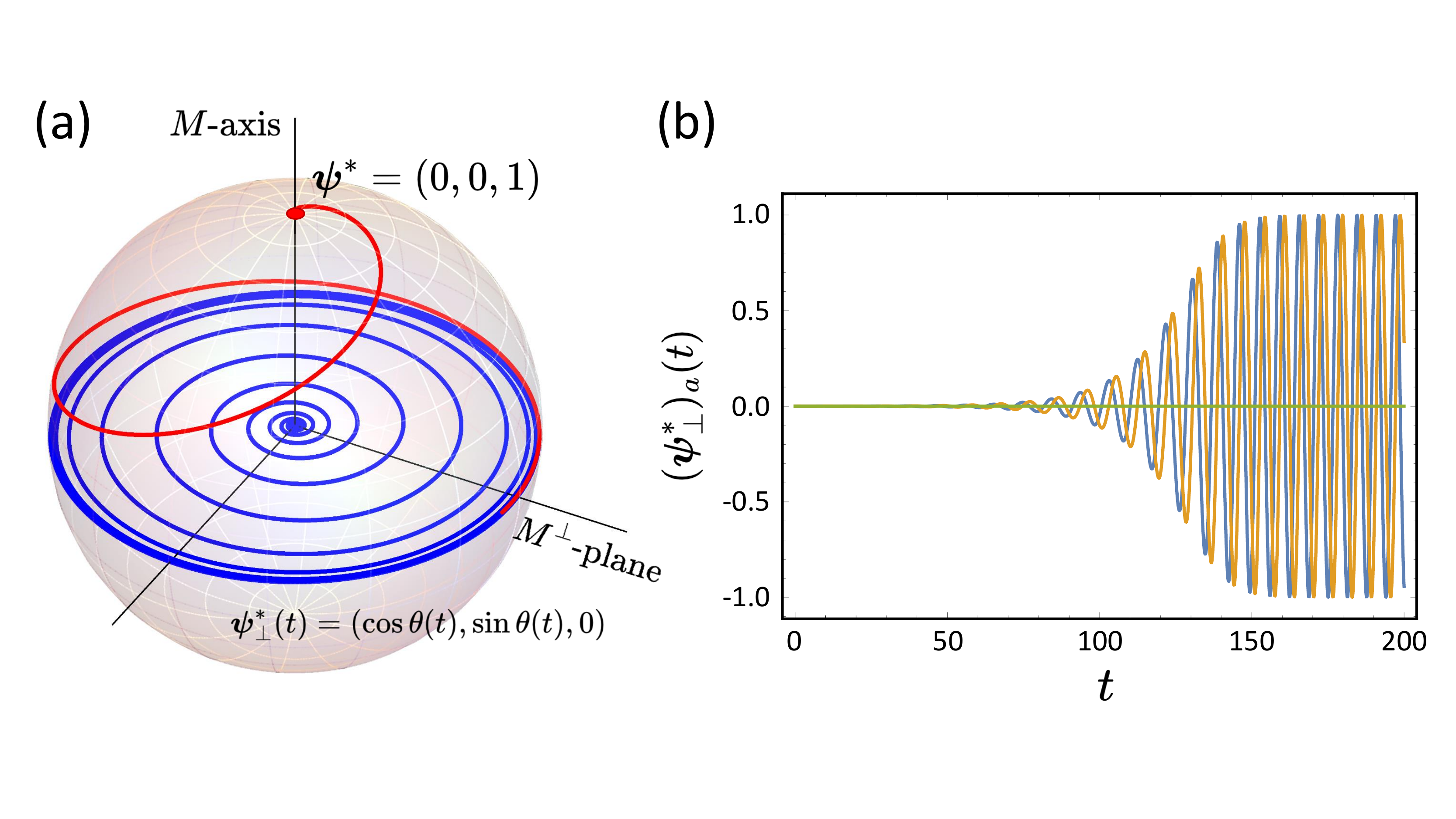}
\caption{(a) A time-parametric plot shows a trajectory initiated close to the origin on the $M^\perp$-plane (blue). This trajectory gradually converges towards a limit-cycle solution on the unit circle within the $M^\perp$-plane, indicating a synchronized solution. If a slight perturbation is introduced along the $M$-axis, the trajectory is abruptly redirected towards the north pole (red). (b) The plot illustrates the time evolution of the components of the WS variable $\bm{\psi}^*_\perp(t)$. The rotation occurs at a tangential speed of $\sin\alpha$. Blue: $(\bm{\psi}_\perp^*)_1$, Orange: $(\bm{\psi}_\perp^*)_2$ and Green: $(\bm{\psi}_\perp^*)_3=0$. } 
\label{Fig:Benjamin-Feir_3D}
\end{figure}

In this section, we provide an example of stability analysis for the synchronized state in $\mathbb{R}^3$. First, consider the synchronized state on the $M$-axis $\bm{\psi}^* = (0,...,0,\pm 1)^\top$, i.e., the north and the south poles, respectively. In fact, the dynamics asymptotically approaches this solution as $t \rightarrow \infty$ since $h(|\bm{\psi}|^2,M) = \frac{|\bm{\psi}|-(1-|\bm{\psi}|^2)\tanh^{-1}|\bm{\psi}|}{|\bm{\psi}|^3}$ for $M=3$ and $\tanh^{-1}(1) = \infty$ while $h(|\bm{\psi}|^2,M) \rightarrow 1$ as $|\bm{\psi}| \rightarrow 1^-$. To study the stability of this solution, let $\bm{\psi}^* = (0,0,1-\delta)^\top$. Then, the trajectory approaches this fixed point solution on the north pole: $\lim_{\delta \rightarrow 0^+} F(\bm{\psi}^*) = (0,0,0)^\top$ where $F(\bm{\psi}):=-\bm{\psi}\bm{g}^\dag \bm{\psi}+\bm{g}$. The Jacobian matrix evaluated at this solution is given by
\begin{flalign}
\lim_{\delta \rightarrow 0^+} J \bigg|_{\bm{\psi}^*} = \begin{pmatrix}
 -1+\cos\alpha & -\sin\alpha  & 0  \\
\sin\alpha & -1+\cos\alpha & 0  \\
0 & 0 & -2 
\end{pmatrix}
\end{flalign} and its eigenvalues are $\lambda_1 = -2, \lambda_{\pm} = -1 + e^{\pm i \alpha}$. Hence, the synchronized solution on the north pole is always a stable solution regardless of $\alpha$. If the trajectory starts slightly outside the $M^\perp$-plane, the solution asymptotically approaches the north pole (or the south pole, depending on whether it initiates above or below the plane.) as time goes on (Fig.~\ref{Fig:Benjamin-Feir_3D} (a)).

Next, we find a trajectory starting on the $M^\perp$-plane for $\alpha <\frac{\pi}{2}$ asymptotically approaching a limit-cycle solution rotating around the great circle on $x_1x_2$-plane (Fig.~\ref{Fig:Benjamin-Feir_3D} (a-b)). The limit-cycle solution can also be interpreted as the synchronized state, as the magnitude of the WS variable is equal to unity. First, letting $\bm{\psi}^*_\perp(t) = \varepsilon (\cos\theta(t),\sin\theta(t),0)^\top$ gives $\lim_{\varepsilon \rightarrow 1^-}F(\bm{\psi}^*_\perp) = \sin\alpha (-\sin\theta(t),\cos\theta(t),0 )^\top$. This demonstrates that the trajectory asymptotically converges to the synchronized limit-cycle trajectory, which rotates counterclockwise along the unit circle with a tangential speed of $\sin\alpha$. Due to its rotational symmetry, we can set the synchronized solution as a fixed point on the unit circle like $\bm{\psi}^*_\perp = \lim_{\varepsilon \rightarrow 1^-}\varepsilon (1,0,0)^\top$. Subsequently, the eigenvalue of the Jacobian matrix, evaluated at this solution, with the corresponding eigendirection along the $M$-axis, is given by $\lambda_3=1-\cos\alpha$. Thus, the synchronized limit-cycle solution is always unstable along the $M$-axis, while it remains stable on the $M^\perp$-plane for $\alpha < \alpha_\text{BF}$.

\section{\label{append:derivation2}Derivation of Eq.~(\ref{eq:magnitude_dynamics_real})}

To obtain Eq.~(\ref{eq:magnitude_dynamics_real}), we consider
\begin{flalign}
\partial_t |\bm{\psi}_1|^2 &= \partial_t(\bm{\psi}_1^\dag \bm{\psi}_1) \notag \\
&=-\mu h_1 \big( \bm{\psi}_1^\dag \bm{K} \bm{\psi}_1 \bm{\psi}_1^\dag -\bm{\psi}_1^\dag \bm{K}^\dag  \big) \bm{\psi}_1 -\nu h_2\big( \bm{\psi}_1^\dag \bm{K}\bm{\psi}_2 \bm{\psi}_1^\dag -\bm{\psi}_2^\dag \bm{K}^\dag \big)\bm{\psi}_1 \notag \\
&~~ ~~-\mu h_1 \bm{\psi}_1^\dag \big( \bm{\psi}_1 \bm{\psi}_1^\dag \bm{K}^\dag \bm{\psi}_1 - \bm{K}\bm{\psi}_1\big) -\nu h_2 \bm{\psi}_1^\dag \big( \bm{\psi}_1 \bm{\psi}_2^\dag \bm{K}^\dag \bm{\psi}_1-\bm{K}\bm{\psi}_2 \big) \notag \\
&= 2(1-|\bm{\psi}_1|^2) \bigg( \mu h_1 \braketmatrix{\bm{\psi}_1}{\frac{\bm{K}+\bm{K}^\dag}{2}}{\bm{\psi}_1} +\nu h_2 \braketmatrix{\bm{\psi}_1}{\bm{K}}{\bm{\psi}_2} \bigg) \notag \\
\partial_t |\bm{\psi}_2|^2 &= \partial_t(\bm{\psi}_2^\dag \bm{\psi}_2) \notag \\
&=2(1-|\bm{\psi}_2|^2) \bigg( \mu h_2 \braketmatrix{\bm{\psi}_2}{\frac{\bm{K}+\bm{K}^\dag}{2}}{\bm{\psi}_2} +\nu h_1 \braketmatrix{\bm{\psi}_1}{\bm{K}^\dag}{\bm{\psi}_2} \bigg) \notag \\
\partial_t \braketmatrix{\bm{\psi}_1}{\bm{K}}{\bm{\psi}_2} &= -\mu h_2 \big( |\bm{\psi}_1|^2\cos\alpha \braketmatrix{\bm{\psi}_1}{\bm{K}}{\bm{\psi}_2} -\braket{\bm{\psi}_1}{\bm{\psi}_2} \big) -\nu h_2\big (\braketmatrix{\bm{\psi}_1}{\bm{K}}{\bm{\psi}_2}^2-|\bm{\psi}_2|^2 \big) \notag \\
&~~ ~~-\mu h_2 \big( |\bm{\psi}_2|^2 \cos\alpha \braketmatrix{\bm{\psi}_1}{\bm{K}}{\bm{\psi}_2}-\braketmatrix{\bm{\psi}_1}{\bm{K}^2}{\bm{\psi}_2}\big) \notag \\
&~~ ~~-\nu h_1 \big( \braketmatrix{\bm{\psi}_1}{\bm{K}}{\bm{\psi}_2} \braketmatrix{\bm{\psi}_1}{\bm{K}^\dag}{\bm{\psi}_2} -|\bm{\psi}_1|^2\cos2\alpha \big) \label{eq:derv1}
\end{flalign} where $h_a := h(|\bm{\psi}_a|^2,M)$ for $a=1,2$. Then, we consider the cross term with the coupling matrix as a variable, i.e., $\xi:=\braketmatrix{\bm{\psi}_1}{\bm{K}}{\bm{\psi}_2} = |\bm{\psi}_1| |\bm{\psi}_2|\cos\theta$ leads to
\begin{flalign}
\braket{\bm{\psi}_1}{\bm{\psi}_2} &= |\bm{\psi}_1| |\bm{\psi}_2|\cos(\theta-\alpha) = \xi\cos\alpha +\sin\alpha \sqrt{|\bm{\psi}_1|^2|\bm{\psi}_2|^2-\xi^2} \notag \\
\braketmatrix{\bm{\psi}_1}{\bm{K}^\dag}{\bm{\psi}_2} &=|\bm{\psi}_1| |\bm{\psi}_2|\cos(\theta-2\alpha) = \xi\cos2\alpha +\sin2\alpha \sqrt{|\bm{\psi}_1|^2|\bm{\psi}_2|^2-\xi^2} \notag \\
\braketmatrix{\bm{\psi}_1}{\bm{K}^2}{\bm{\psi}_2} &=|\bm{\psi}_1| |\bm{\psi}_2|\cos(\theta+\alpha) = \xi \cos\alpha-\sin\alpha \sqrt{|\bm{\psi}_1|^2|\bm{\psi}_2|^2-\xi^2}.
\label{eq:derv2}
\end{flalign} Note that for even $M$, we can easily find $\braketmatrix{\bm{\psi}_a}{\frac{\bm{K}+\bm{K}^\dag}{2}}{\bm{\psi}_a} = \cos\alpha |\bm{\psi}_a|^2$. However, for odd $M$, we find that chimera states only live on $M^\perp$-plane and hence we assume  $\braketmatrix{\bm{\psi}_a}{\frac{\bm{K}+\bm{K}^\dag}{2}}{\bm{\psi}_a} = \cos\alpha |\bm{\psi}_a|^2$ for $a=1,2$. Finally, plugging Eq.~(\ref{eq:derv2}) into Eq.~(\ref{eq:derv1}), we obtain Eq.~(\ref{eq:magnitude_dynamics_real}) in Sec.~\ref{sec:real_system}.


\section*{Acknowledgments}
This work has been supported by the Deutsche Forschungsgemeinschaft (DFG project KR1189/18-2). The authors would like to thank Julius Fischbach for a fruitful discussion.

\vskip 1cm 
\section*{References}
\bibliographystyle{unsrt}
\bibliography{ref.bib}

\end{document}